\documentstyle[12pt,aasms]{article}

\newcommand{\gsim}{\lower .5ex\hbox{$\buildrel > \over {\sim}$}}
\newcommand{\lsim}{\lower .5ex\hbox{$\buildrel < \over {\sim}$}}
\newcommand{\mdot}{{\rm M}_\odot}
\newcommand{\mz}{M_{_{\rm ZAMS}}}
\newcommand{\etal}{{\it et~al.}}
\newcommand{\refmark}[2]{#1}
\newcommand{\bfdot}[1]{\bf\dot{\rm #1}\rm}

\begin{document}

\title{On the Nature of Core-Collapse Supernova Explosions}
\author{Adam Burrows and John Hayes}
\affil{Departments of Physics and Astronomy, University of Arizona, Tucson, AZ
85721}

\and

\author{Bruce A.~Fryxell}
\affil{Goddard Space Flight Center, NASA, Greenbelt, MD  20771}

\begin{abstract}
We investigate in this paper the core-collapse supernova explosion mechanism
in both one and two dimensions. With a radiation/hydrodynamic code based upon
the PPM algorithm, we verify the usefulness of neutrino-driven
overturn (``convection'') between the shock and the neutrinosphere in igniting
the supernova explosion.
The 2-D simulation of the core of a 15$\mdot$ star that we present here
indicates
that the breaking of spherical symmetry may be central
to the explosion itself and that a multitude of bent and broken fingers is a
common
feature of the ejecta.
As in one-dimension, the explosion seems to
be a
mathematically {\bf critical} phenomenon, evolving from a steady-state to
explosion
after a critical mass accretion rate through the stalled shock has been
reached. In the 2-D simulation we show here, the pre-explosion convective phase
lasted $\sim$30 overturns ($\sim$100 milliseconds) before exploding.
The pre-explosion steady-state in 2-D is similar to
that achieved in 1-D, but, in 2-D, due to the higher dwell time of matter in
the overturning
region, the average entropy achieved behind the stalled shock
is larger. In addition, the entropy gradient in the convecting region is
flatter. These effects,
together with the dynamical pressure of the buoyant plumes, serve to increase
the steady-state shock radius ($R_s$) over its value in 1-D by 30\%--100\%.
A large $R_s$ enlarges the
volume of the gain region, puts shocked matter lower in the gravitational
potential well, and lowers the accretion ram pressure at the shock for a given
$\bfdot M$. The critical condition for explosion is thereby relaxed.
Since the ``escape'' temperature ($T_{\rm esc}$)
decreases with radius faster than the actual matter temperature $(T)$ behind
the shock, a larger $R_s$ puts a larger fraction of the shocked material above
its local escape temperature. $T>T_{\rm esc}$ is the condition for
a thermally-driven corona to lift off of a star. In one, two, or three
dimensions, since supernovae are driven by neutrino heating, they are coronal
phenomena, akin to winds, though initially bounded by an accretion
tamp. Neutrino radiation pressure is unimportant.

We find that large and small eddies coexist,
both before and after explosion.
In the unstable region before explosion, columnar downflows are
quasi-periodically formed and break up. These plumes excite non-linear
internal g-modes that feed back onto the convection and cause the plumes to
meander over the neutrinosphere. The radial neutrino flux fluctuates with
angle and time in response to the anisotropic mass flux onto the
neutrinospheres by as much as a factor of three. The boiling motion of the
unstable region interior to the shock is epitomized by neutrino-heated bubbles
that rise and collide episodically with the shock, whose radius oscillates in
angle and time by as much as 30\%. The angle-averaged neutrino luminosities
vary by as much as 60\% and decrease by a factor of two right after the
explosion in a characteristic way.

The region interior to the neutrinosphere has weakly unstable lepton and
entropy gradients that drive persistent convective motions after core
bounce. However, the effects of this convection on the driving neutrino
luminosities seem dwarfed by the effects of convective dredge up and by the
wildly varying accretion component.

We see no evidence of a ``building'' or accumulation of energy before
explosion, save in
the kinetic energy, in response to the decaying accretion ram.  In fact, the
total energy in the overturning
region {\it decreases} steadily before explosion.
In addition, we have noted a non-trivial dependence
on the neutrino transport algorithm.  We would eschew terms
such as ``robust'' when referring to the effect of convection on the
outcome of collapse.

Neutrino energy is pumped into the supernova during the shock's propagation
through the inner many thousands of kilometers and not ``instantaneously.''
Curiously, just after the explosion is triggered, the matter that will
eventually
be ejected is still {\bf bound}.
In addition, for a given asymptotic explosion energy, the amount of mass
that reaches explosive nucleosynthesis temperatures is less than heretofore
assumed. This may
help to solve the $^{56}$Ni overproduction problem encountered in previous
models of explosive nucleosynthesis.

The high-speed fingers that emerge from our model core seem a natural
explanation for the nickel bullets seen in SN1987A and the shrapnel inferred
in some supernova remnants. In addition, the vigorous convective motions
interior to the shock can impart to the residue recoil velocities and
spins. The magnitudes of the former might be within reach of the observed
pulsar proper motions, but extensive new calculations remain to be done to
verify this. Within 100 milliseconds of the explosion, a strong,
neutrino-driven wind is blowing outward from the protoneutron star that clears
the interior of mass and, while operative, does not allow fallback. At the
base of the rising
explosion plumes (in the early wind), a few high entropy $(\sim60)$ clumps are
ejected, whose subsequent evolution may prove to be of relevance to the
r-process.
\end{abstract}

\section{Introduction}
The supernova problem has been solved many times in the last thirty years, but
never yet for long. Its persistence as a puzzle has many causes: 1)
galactic supernovae are rare ($\sim$1/30--100 years), 2) expertise in a broad
range of scientific and numerical subdisciplines seems to be required, 3) the
launching of the explosion is obscured by a massive envelope impenetrable by
photons, and 4) extragalactic supernovae have not until recently been observed
in sufficient detail to provide theorists with hard constraints. Furthermore,
the community has been divided into those emphasizing as primarily
important either nuclear physics, progenitor models, neutrino interactions,
multi-dimensional effects, numerical rigor, or exotic processes. This
fragmentation may merely reflect the richness of a subject that involves most
of twentieth-century physics. However, its complexity and highly nonlinear
nature have left supernova theory vulnerable in the past to simplistic
solutions that are only slowly refuted. Nevertheless,  there has been much
progress
in the last few years in distinguishing the essential from the non-essential
ingredients in a detailed
supernova simulation, in discovering what does not work, and in establishing
what will be required to satisfy the growing number of observational
constraints. In particular, an understanding of hydrodynamic instabilities and
overturn before, during, and after core collapse now seems to be central to
the resolution of some or all of the supernova problem. This paper is the
first in a new series of papers on the mechanism of supernova explosions. We
present new one- and two-dimensional radiation/hydrodynamic simulations of the
collapse, bounce, and explosion of the cores of massive stars, with special
emphasis on multi-dimensional effects.  In \S II, we describe the numerical
scheme we have developed to
perform supernova simulations in one and greater dimensions. In \S III , we
discuss, in the
context of new 1-D hydrodynamic simulations of stellar collapse, bounce, shock
formation, and accretion, the demise of the prompt mechanism and quantities
relevant in the study of the supernova mechanism. The roles and
characteristics of the progenitor models are discussed in \S IV. This is
followed in \S V by a discussion of the neutrino-driven mechanism in
one-dimension.
The general
physics of convective instability is described in \S VI, in which we also
survey the status of multi-dimensional supernova work. In \S VII, our new
two-dimensional supernova explosion calculations are depicted, described, and
analyzed. This is followed in \S VIII with
a pr\'ecis of our conclusions concerning multi-dimensional supernovae.

\section{The Numerical Scheme}
The multi-dimensional computer code we have used to simulate supernova
explosions is a robust tool that can follow in a self-consistent fashion
evolution from stellar collapse, to bounce, to explosion, through to the
propagation of the blast out into the progenitor. To achieve flexibility and
some speed, we have made many approximations in the neutrino transport
module. In particular, we solve the neutrino {\it diffusion} equations, not
the Boltzmann or transport equations, we have simplified the neutrino cross
sections and sources, and we have dropped all relativistic terms. In addition,
the gravity is Newtonian and only the monopole term in the potential is
retained. The basic hydrodynamics code that we have embellished is
PROMETHEUS (\cite{fry89}, \refmark{1991}{fry91}), itself a
realization of the \underbar{P}iecewise-\underbar{P}arabolic-\underbar{M}ethod
(PPM) of \refmark{Colella and Woodward (1984)}{col84}. This hydro portion of
the code
is automatically conservative, explicit, Eulerian, and second-order
accurate in space and time (error of $O(\Delta x)^2$ and $O(\Delta t)^2$),
with a Riemann solver that enables us to
resolve the shock position to within two zones. The diffusive transport is
done implicitly to
avoid severe time step limitations in the core, where neutrino processes
achieve thermal and chemical equilibrium very quickly during post-bounce
phases.

An Eulerian collapse code has intrinsic virtues and vices. Since the grid is
fixed in space and not in mass, winds and explosions can be more realistically
handled than in Lagrangian codes. This is particularly important for supernova
simulations. However, since the zones do not ride on the collapsing mass, one
must pre-zone the core quite accurately to resolve the violence of bounce and
shock formation. Typically, we have 100--400 radial zones interior to 100
kilometers
during these dynamical phases, out of a total of 600--700 zones that span the
entire inner 4500 kilometers (for models w$\ast$n or w$\ast$t). We have
installed a
rezoning feature that, while not ``dynamical,'' allows us to redistribute
zones at various phases during a calculation, should we so desire. This
1-D$\to$1-D mapper is augmented by a 1-D$\to$2-D mapper that allows us to
follow
spherical (1-D) phases in 1-D and map to a 2-D calculation just before
multi-dimensional effects are expected. In this way, we do not waste CPU
time. We
can also cut out an inner sphere in the middle of a calculation and replace it
with a new inner boundary at the sphere's outer surface that automatically
uses the excised zones' densities, pressures, energies, and fluxes for the new
inner boundary. More powerfully, we can do the inner core's hydrodynamics in
one dimension, while continuing to follow the rest of the star in 2-D. As long
as the inner core is spherical, this ``restricted-2-D'' allows us to avoid
severe CFL restrictions, due to the convergence of the angular zones at the
stellar center, while still following the core hydro and transport. A full 2-D
run typically requires 50 trillion floating point operations. On the C90 at
Pittsburgh,
we achieve speeds of about 300 MFlops. The timestep
during most of a 2-D calculation is 1--3 microseconds.

The nuclear equation of state (EOS) we have employed for the calculations
discussed in this paper is similar to that used by \refmark{Burrows and
Lattimer (1985)}{bl85}. The only difference is that in the current version
the symmetry energies we employ are those of Lattimer and Swesty (1991). The
EOS assumes NSE
and a single representative heavy nucleus,
includes alphas, neutrons, protons, photons, pairs, and neutrinos (only in the
opaque regions), and is fully vectorized for optimal operation on CRAY
architectures. Recently, we have tabulated for hydrodynamic use the full
\refmark{Lattimer and Swesty (1991)}{lat91} nuclear equation of state, which
after
sufficient speed and accuracy are achieved, we plan to use in future
calculations. The differences between the two EOS's are slight, except near
the phase transition at $\sim0.5\rho_{{\rm nuc}}$ and at low densities $(\le
10^8
{\rm\ gm/cm}^3)$ and temperatures $(<0.7\ {\rm MeV})$, where our alpha
fractions
are a bit too high.

The basic transport scheme is as described in \refmark{Burrows and Lattimer
(1986)}{bl86} and
\refmark{Burrows and Fryxell (1993)}{bf93}. Radial transport along  different
angular rays is followed
independently, but angular transport is
suppressed, thereby suppressing various neutrino viscosity effects that are
ignored in the basic hydro structure of PPM anyway. The transport is coupled
to the hydro in operator-split fashion. Three neutrino fluids,
$\nu_e$'s, $\overline{\nu}_e$'s, and ``$\nu_\mu$'''s $(\nu_\mu,
\overline{\nu}_\mu, \nu_\tau, \overline{\nu}_\tau$'s) are followed in the
diffusive approximation. In the semi-transparent regions $(\tau_\nu<0.5)$, the
rates of neutrino absorption and emission are assumed to be of the following
simple forms.

For $\nu_e(n,p)e^-$ and $\overline{\nu}_e(p,n)e^+$,
$$
\bfdot\epsilon_+=2.0\times 10^{18} T_{\nu_e}^6f\left({F_5(\eta_{\nu_e})\over
F_5(0)} Y_n+\left({T_{\overline\nu_e}\over T_{\nu_e}}\right)^6
{F_5(\eta_{\overline\nu_e})\over F_5(0)} Y_p\right) {{\rm ergs}\over{\rm
gm}\cdot{\rm s}}\eqno(1)$$
and
$$\bfdot\epsilon_-=2.0\times 10^{18}T_e^6\left[{F_5(\eta_e)\over F_5(0)}
Y_p+{F_5(-\eta_e)\over F_5(0)} Y_n\right] {{\rm ergs} \over {\rm gm}\cdot {\rm
s}},\eqno (2)$$
where $T_{\nu_e}$ and $T_{\overline\nu_e}$ are the electron neutrino and
antineutrino temperatures at their respective neutrinospheres, $\eta_{\nu_e}$
and $\eta_{\overline\nu_e}$ are the assumed spectral pinch factors
(\cite{myr90}), $\eta_e$ is ${\mu_e\over T_e}$, and $T_e$
is the matter temperature. Guided by the
work of Myra and Burrows (1990), we assume in these calculations that the
emergent
$\nu_e$ and $\overline{\nu}_e$ spectra are Fermi-Dirac in shape, with a default
$\eta$ of
2 and a temperature equal to the matter temperature at decoupling.
The $F_n(x)$'s are the standard relativistic Fermi
integrals. $f$ is the spherical dilution factor that has the correct values at
the neutrinospheres $(R_\nu)$ and for $r> > R_\nu$ and which we set equal to
${1\over 2}(1+({R_\nu\over r})^2)(1-\sqrt{1-({R_\nu\over r})^2})$. The pair
loss rate is taken to be
$$
\bfdot\epsilon_p=1.5\times 10^{25}{T_e^9\over\rho}\left[{F_4(\eta_e)
F_3(-\eta_e) +F_4(-\eta_e) F_3(\eta_e)\over 2F_4(0) F_3(0)}\right] {{\rm
ergs}\over [{\rm gm}\cdot {\rm s}]}.\eqno (3)$$
As in \refmark{Schinder and Shapiro (1982)}{sch82}, neutrino-electron
scattering
is
handled for each $\nu_i-e^\pm$ combination as a Compton
scattering problem:
$$\bfdot\epsilon_{\nu_ie}={1.64\times 10^{24}\over\rho}\Lambda_i
\left(1+{\eta_e\over 4}\right) {F_2(\eta_e) F_4(\eta_{\nu_i})\over F^2_3(0)}
(T_{\nu_i}-T_e) T^4_{\nu_i} T^4_e f {{\rm ergs}\over{\rm gm}\cdot{\rm s}},
\eqno(4)$$
where $\Lambda_i$ is a function only of the Weinberg angle.
The electron {\it number} sources and sinks are derived using the same
approach that gave eqs.\ (2) and (3), with the appropriate changes of sign,
Fermi indices, and powers of $T$. Note that $\bfdot\epsilon_p$ is greater than
$\bfdot\epsilon_+$ only for entropies above $\sim 70$.

The neutrino-matter cross sections used in the diffusive regime are those due
to the standard neutral- and charged-current neutrino-nucleon interactions,
with no corrections for ion-ion correlation or screening
(\cite{lb91}). Neutrino-electron scattering is ignored in the opacity and
the matter and neutrino temperatures are assumed equal when
$\tau_\nu>2/3$. Levermore-Pomraning flux limiters are used.
In addition, our Eulerian scheme does not capture the rapid break-out
phenomenon very well and our neutrino luminosities between bounce and 10
milliseconds
are not to be trusted.  However, after break-out, our energy luminosities are
competitive.
Though this
approach is highly simplified and a multi-group treatment would be preferable,
such a more detailed scheme would multiply our CPU requirements by more than a
factor of twenty.
Nevertheless, our general results in 1-D are similar to
those obtained by others who employ more precise neutrino transport schemes.

\vfill\eject

\section{The Fall of the Prompt Mechanism}
This section will  not be a review of the details of stellar collapse, but a
discussion and listing of some of its essentials. For a broader overview, the
reader is referred to \refmark{Burrows (1990)}{bur90}, \refmark{Bruenn (1985,
1989a,b)}{bru85,br89a,br89b},
\refmark{Woosley and Weaver (1986)}{ww86}, and \refmark{Bethe (1988)}{bet88}.

At collapse, depending on the specific calculation and the progenitor mass
 ($M_{_{\rm ZAMS}}\gsim 8\ \mdot$), the
central density $(\rho_c)$, temperature $(T_c)$ and entropy are $4\times
10^9-10^{10}$ gm/cm$^3$, $\sim 0.5$ MeV, and $0.5-1.2$ per baryon per
Boltzmann's constant, respectively. Due to electron capture during and after
core carbon and oxygen burning, the central electron fractions $(Y_e)$ hover
around 0.43 and the electron chemical potential $(\mu_e)$ is $8-10$ MeV
(\cite{auf94}; \cite{nom85}).
The ``Chandrasekhar'' mass of the iron $(\mz>10\ \mdot)$ or ONeMg
$(8\ \mdot<\mz\lsim\ 10\ \mdot)$ cores ranges between 1.2 $\mdot$ and $\sim
2.0\
\mdot$ and the
density structures show a correlated variation that determines the outcome of
collapse (see \S IV). The more compact configurations collapse to nuclear
densities more quickly than those that are more extended. To illustrate
stellar collapse, we have started with the 15 $\mdot$ (s15s7b) and 20 $\mdot$
(s20s7b) core models of \refmark{Weaver and Woosley (1995)}{ww95}. Figures 1a
and
1b depict the evolution in 1-D of the mass density profiles versus interior
mass from
when a velocity near minus $10^8$ cm/s is first approached (``a'') to 50--60
milliseconds after core bounce (``d''). Seven hundred radial zones were
used. To ensure absolute consistency when comparing one-dimensional and
two-dimensional calculations, we always used exactly the same code (see \S
II). Model w15t takes 209 milliseconds to bounce at $\rho_c\sim3.8\times
10^{14}$ gm/cm$^3$ and model w20t takes $\sim547$ milliseconds, but in both
$\rho_c$ has climbed 4.5--5 orders of magnitude. The evolution of the central
density for various 1-D models mentioned in this study is summarized in Table
1.
The central density takes
145--167 milliseconds to go from $10^{10}$ to $10^{11}$ gm/cm$^3$, then 20--25
milliseconds to achieve $10^{12}$ gm/cm$^3$, then $\sim 5$ milliseconds to
reach $10^{13}$ gm/cm$^3$, and finally only $\sim1.5$ milliseconds to attain
$10^{14}$ gm/cm$^3$. Within one millisecond of bounce, a strong sound wave
that steepens into a shock wave near 0.6--0.8 $\mdot$ is generated and
penetrates the electron neutrinosphere near $10^{11}$ gm/cm$^3$ and
0.9--1.0 $\mdot$. However, within ten milliseconds of bounce, the shock stalls
near 90--110 kilometers and 1.2 $\mdot$ due to neutrino losses at shock
break-out and the dissociation of nuclei, the former being the more
important. The bounce-shock fails to explode. Figures 2a and 2b depict the
evolution of velocity versus interior mass and nicely illustrate the growth
of the shock and its stall into accretion. Similarly, the evolution of the
entropy and $Y_e$ profiles to $\sim60$ milliseconds after bounce is depicted
in Figures 3a, 3b, 4a, and 4b. A few generic features deserve mention. The
entropy behind the shock first peaks near 0.9--0.95 $\mdot$ (and near shock
break-out of the electron neutrinospheres) at a value of 8--10. As the shock
stalls, it leaves behind a negative entropy gradient that is unstable to
overturn, but is smoothed by neutrino heat transport within $\sim$15
milliseconds. However, though the entropy at $\sim 0.9\ \mdot$
decreases within tens of milliseconds to $\sim5$, a region of negative entropy
gradient between $\sim0.9\ \mdot$ and 1.2--1.4 $\mdot$
develops in the same region that the neutronization trough (\cite{bm83})
appears. In this mass range,
$Y_e$ and entropy gradients that would be Rayleigh-Taylor {\it unstable} in
two or three dimensions are formed, in equilibrium with the neutrino
diffusion of heat and leptons. Just behind the shock, electron capture on
newly-liberated protons is rapid and can leave the matter with $Y_e$'s less
than 0.1. In these calculations, the central trapped lepton fraction $(Y_e)$
is 0.35--0.38, a bit higher than the preferred values ($\sim$0.33--0.35,
\cite{bru85}, \refmark{1992}{bru92}). If our electron capture and
neutrino transport algorithm is modified to allow more capture on infall and
delayed trapping, the smaller consequent $Y_\ell$ ($\sim$0.34) results in a
weaker bounce
and shifts the first entropy peak to a value of $\sim$7 near 0.8 $\mdot$. In
this case, the diffusion of heat does not smooth the negative entropy gradient
imposed on the matter as the shock stalls. Thus, we see that the trapped
lepton fraction has a quantitative effect on where and when matter is unstable
to overturning motions, but that unstable lepton and entropy gradients are
always obtained early and maintained. This is important in the
subsequent development of hydrodynamic instabilities. The sharp entropy spike
seen just behind the shock in Figures 3 is a consequence of neutrino heating
and has been seen by others (e.g. Bruenn 1992). In our calculations, the
position of the shock
is resolved to better than four kilometers. However, there are a number of
vices in our approach that should be mentioned (see also \S II). Capture on
heavies is not included, electrons are assumed to be relativistic, general
relativity is ignored, the phase transition near nuclear density is not done
consistently, and neutrino transport is done in the diffusive
approximation. Nevertheless, the generic character of the
more detailed calculations of \refmark{Bruenn (1992)}{bru92}, \refmark{Wilson
(1985)}{wil85},
\refmark{Wilson and Mayle (1993)}{wil93}, and \refmark{Swesty \etal\
(1994)}{swe94} is reproduced. The luminosities, shock radii, velocities, and
timescales,
etc. are not precisely the same as obtained in those other papers. In
particular, the
advection terms in our transport module are not robust enough to accurately
handle
the rapid and violent break-out phenomenon.  As a result, our break-out
luminosities, in particular
our anti-electron neutrino luminosities, are not correct during the first ten
milliseconds after shock formation.  Within twenty milliseconds of break-out,
the scheme is back on track
and any differences between our luminosities and those of others at this time
are
predominantly due to the different
opacities, neutrino processes, and progenitors employed by the various groups.
In particular, the opacities that we use result in slightly larger (but not
extraordinarily so)
core luminosities during the
delay phase of the evolution.
It is important to note that the
differences after break-out between our results and those of other theorists
are smaller than
the range of results still possible due to the ambiguities in the nuclear
symmetry energies, the nuclear incompressibility, and the neutrino opacities
(\cite{swe94}; \cite{bru94}).

The positions of the shock and the various neutrinospheres (defined by
$\tau=2/3$) are depicted in Figures 5a and 5b for models w15t
and w20t, respectively.
For comparison, the angle-averaged shock radius
versus time for the 2-D {\bf star} calculation is depicted in Figure 5c and
compared
to the corresponding curve for its 1-D analog.  This important plot will be
discussed in \S VII.
In the 1-D calculations,
the shock settles to near 80 kilometers and sinks at a rate of 0.3--0.4
kilometers per millisecond. The neutrinospheres are 20--30 kilometers interior
to the shock, between 35 and 60 kilometers from the center, and sink at an
average initial rate of $\sim$0.4 kilometers per millisecond. The differences
between the characteristics of the 15 $\mdot$ and 20 $\mdot$ models are not as
important as their similarities.
Figures 6a and 6b show the evolution of the neutrino
luminosities 60 milliseconds before and after bounce for 1-D models w15t and
w20t.

As stated above, the radiation module of the code does not handle the shock
break-out
phenomenon particularly well
since this phase is fundamentally dynamical and rapid. As a consequence, the
ringing seen in Figures 6 (with a period of $\sim$5 milliseconds) is real, but
its amplitudes may be smaller and the ratio of the $L_{\overline\nu_e}$ to
$L_{\nu_e}$ within $\sim$10 milliseconds of break-out may be
corrupted. In addition, our breakout $L_{\nu_{\mu}}$'s are probably too large.
Nevertheless, the rapid rise time of $L_{\nu_\mu}$ and
$L_{\overline\nu_e}$ and the subsequent suppression of $L_{\overline\nu_e}$
relative to $L_{\nu_e}$ due to the preponderance of electron neutrinos near
the neutrinospheres are generic. The ringing is damped out after
3--4 cycles. The electron neutrino luminosity has a pre-bounce rise time of
10--16 milliseconds that reflects the nuclear symmetry energy employed
(\cite{swe94}). Due to higher accretion rates, the
luminosities in the 20 $\mdot$ model are larger than in the 15 $\mdot$ model,
and
they decay more slowly. A slow convergence of $L_{\nu_e}$ and
$L_{\overline\nu_e}$ (not seen in Figures 6) is a consequence of the high
post-shock capture rates for models with small equilibrium shock radii
($R_s\sim$80 km, as opposed to 150 km).

Figures 1--6 demonstrate, as has been concluded many times by others
(e.g. \cite{maz82}; \cite{bl85};
\cite{bru85}; \cite{br89a}, \cite{bru92};
\cite{wil85}; \cite{bar90}) that the
direct mechanism of supernova
explosions (\cite{col60}) does not work. The pressure
deficits due both to electron capture and the radiation of neutrinos of {\it
all} species and to the photo-dissociation of infalling nuclei can not be
overcome. The shock must stall within $\sim$10 milliseconds of its birth. This
conclusion  does not depend on the incompressibility of nuclear matter, the
trapped lepton fraction, or general relativity (\cite{swe94};
\cite{bl85}; \cite{bru85}). Another
simple
way to see why the shock stalls is to compare the hydrodynamic power
($L_{_H}=4\pi r^2 Pv$ or $4\pi r^2 ({1\over 2})\rho v^3$) to the sum of the
neutrino luminosities $(L_{_T})$ and $L_{_D}=\bfdot{M_s} \epsilon_d$, where
$\epsilon_d$ is the energy
required to dissociate a unit mass of ``iron'' ($\cong$ 8.8 MeV/b) and
$\bfdot{M_s}$ is the rate at which the shock encompasses mass (see below).
While
$L_{_H}$ near break-out is a few times $10^{53}$ ergs/s, $L_{_T}$ hovers near
$10^{54}$ ergs/s and $L_{_D}$ is also a few times $10^{53}$ ergs/s. Crudely,
the shock stalls because $L_{_H}\lsim L_{_T}+L_{_D}$. Various codes and
progenitors will result in
different entropy and lepton profiles and stalled shock positions, but the
fundamental outcome is the same. There has been considerable effort to prove
otherwise to no avail. Even if the bounce shock were to succeed in a situation
with ``all the physics'' ostensibly included, such an explosion would leave
behind too small a neutron star ($\sim$ 1.1 $\mdot$ gravitational) and eject
too much neutron-rich material $(\sim 0.2-0.3\ \mdot)$ (\cite{thn90}).

There is nothing particularly esoteric or controversial about this
conclusion. The explosion must occur after the shock stalls and is a function
of the behavior of the quasi-hydrostatic protoneutron star, bounded by an
accretion shock. The heating of the shocked envelope by the neutrinos from the
core has been suggested as the driver of the explosion (\cite{bet85} and \S
V), but it is yet to be shown that explosions with
the required energy can occur after a delay without some sort of convective
instability (\S VI and \S VII).

Before we turn to a short discussion of the progenitor models, some numbers
derived from simulations w15t and w20t will prove useful. Though both the ``15
$\mdot$'' and ``20 $\mdot$'' progenitor cores behave similarly, there are
interesting differences that stem almost exclusively from their core
structures. The 15 $\mdot$ model has an iron core mass of 1.28 $\mdot$ and a
steep density gradient exterior to it that initially follows a
$1/r^{(3.5-3.8)}$  power law. The iron core of the 20 $\mdot$ star is
1.74 $\mdot$ and has a shallower density gradient that follows a $1/r^3$ power
law. The
rate at which mass accumulates interior to the shock is very different for the
two progenitors, each representative of the two basic classes of structures
that modelers have published. Within 50 milliseconds of bounce, while the
shock in model w15t bounds $\sim$1.35 $\mdot$, that in model w20t already
bounds $\sim 1.5\ \mdot$. For model w20t, the mass accretion rate through the
shock $(\bfdot{M_s})$ at
this time is twice that for model w15t. Figure 7 depicts this mass accretion
rate through the stalled shock and the mass interior to
the shock $(M_s)$ for models w15t and w20t during the first 60
milliseconds after bounce. Near break-out, $\bfdot{M_s}$ is near 100 $\mdot$/s,
but
within 10 milliseconds it is below 10 $\mdot$/s for both models. At $\sim$50
milliseconds, $\bfdot{M_s}$ is $\sim5\ \mdot$/s in model w20t. Importantly,
$\bfdot{M_s}$
and $M_s$ after the shock stalls are functions, not of the physics of bounce or
neutrinos, but of the
{\it initial} progenitor structures (and the electron capture algorithm
employed on infall). This is a consequence of the fact that the unshocked
mantle is in supersonic infall and is not causally (sonically) connected to
the core. All else being equal, $\bfdot{M_s}$ and $M_s$ after the shock settles
into accretion
are vital in determining
the outcome of collapse (Burrows and Goshy 1993), the delay to explosion
necessary to prevent the contamination of the ISM with exotic nuclear species,
and whether a neutron star of the requisite gravitational mass is left. For
instance,
if the gravitational mass of the residual neutron star is 1.35 $\mdot$, its
baryon mass must be $\sim 1.5\ \mdot$. Model w15t would take $\sim$500
milliseconds to achieve such an $M_s$, while model w20t would require only 50
milliseconds. This does not include the effect of fall back either early
(within 20 seconds) or late, due to the creation of a reverse shock at the
hydrogen/helium interface after many minutes to hours. However, such
considerations set the timescales for action in these models. In addition, the
times after bounce to accrete the iron core edges if an explosion has not
occurred are 25 and $\sim$600 milliseconds for models w15t and w20t,
respectively. Table 2 shows the edge radii for various 1-D models at various
epochs. The locations of the edges of the iron cores at
bounce are 540 and 1430 kilometers, respectively, and of the silicon outer
edges are
2460 and 3270 kilometers, respectively. During infall, the iron core edges
moved 600 kilometers and 800 kilometers for w15t and w20t, respectively. These
numbers, and numbers like them for other supernova codes and progenitor
models, provide the context in which to diagnose the various explosion
scenarios. Another number of relevance is the mass $(\Delta M)$ between the
electron neutrinosphere and the shock versus time, both before and after
explosion. In our models w15t and w20t, $\Delta M$ at 50 milliseconds after
bounce is only $\sim$0.01 $\mdot$ and is decreasing with a mean life of 25--50
milliseconds. This is rather small and implies (with the considerations of \S
V) that in the 1-D calculations what has been accreted through the shock before
explosion will
remain, to within no more than 0.01$\mdot$.

The code that we have constructed
can also be made to generate a variety of explosions by changing a variety of
parameters. Such artificial explosions  can be followed to very large radii
and can be used to
understand blast propagation through the rest of the progenitor in a natural
and consistent way that partitions the internal and kinetic energies
realistically (\cite{auf94}; \cite{burin}). One such calculation is model w15n
in which we have turned
off all neutrino processes and frozen $Y_e$. The velocity versus radius
profiles for this model are depicted in Figure 8. Here, the shock succeeds and
has reached 4100 kilometers within 180 milliseconds of bounce, despite the
dissociation ``losses.''  Clearly, neutrino losses  change the
outcome in a qualitative way (cf. model w15t). The total kinetic energy of the
blast is
$0.85\times 10^{51}$ ergs at 190 milliseconds, but troughs near $0.3\times
10^{51}$ ergs earlier in the explosion. It grows due to the reassociation of
the heavies and alphas previously broken up by the shock. The evolution of the
kinetic energy with time is depicted in Figure 9. It plateaus off the graph
near $\sim
1\times 10^{51}$ ergs. However, the same sort of calculation with the
20 $\mdot$ model (w20n) did not explode. In that simulation, the shock reached
$\sim$800 kilometers before stalling. In the 20 $\mdot$ model, there was too
much matter at small radii to dissociate and too high a gravitational binding
energy to overcome.

\section{Progenitor Structures}
Despite significant progress in recent years, our understanding of massive
star progenitor structures is still incomplete.
 This is illustrated in Figure 10,
which depicts the iron core masses $(M_{{\rm Fe}})$ recently calculated by
\refmark{Weaver and Woosley (1993)}{wea93} and \refmark{Weaver and Woosley
(1995)}{ww95} under various assumptions
and by \refmark{Nomoto and Hashimoto (1988, NH)}{nom88}. The former now prefer
the ``sb'' series, but there is as yet no sign of convergence between the
groups for a given
ZAMS mass. Near 20 $\mdot$, $M_{{\rm Fe}}$ has ranged 0.6 $\mdot$ during the
1980's
and 1990's, depending on the value of the $^{12}$C$(\alpha,\gamma)^{16}$O
rate, the degree of overshoot, the handling of semi-convection, the degree of
electron capture during burning, etc. (\cite{arn91};
\cite{ww86}; \refmark{NH}{nh}). \refmark{Bazan and Arnett (1994)}{baz94} have
recently shown
that after the onset of core oxygen burning convection is dynamical, the
burning timescales are comparable to the convective overturn times, and the
mixing-length prescription currently employed by the stellar modelers is
suspect. As \refmark{Weaver and Woosley (1993)}{wea93} have shown, the mapping
of
ZAMS mass to $M_{{\rm Fe}}$ and core structure is very sensitive to inputs, in
what is perhaps a chaotic way. In addition, mass loss from the most massive
stars
(not included in the calculations of  Weaver and Woosley 1993,1995)
will change the core mass systematics in as
yet unknown ways (Woosley, Langer, \& Weaver 1994).  In short, we can not yet
say with any certainty
what density profiles at collapse and what $M_{{\rm Fe}}$'s correspond to what
ZAMS mass
progenitors. However, there are certain systematics that bear mentioning. As
Figure 10 shows, there is a trend in $M_{{\rm Fe}}$ with $M_{{\rm ZAMS}}$
that reflects
the higher core entropies expected at a given burning temperature for higher
masses (\refmark{NH}{nom88}). With increasing mass, the
relative carbon yields shrink and the duration and degree of carbon and neon
shell burning decreases. This results in higher core entropies, fatter core
envelopes, and higher effective Chandrasekhar masses. Overshoot,
semi-convection, and a high $^{12}{\rm C}(\alpha,\gamma)^{16}{\rm O}$ rate
generally do the same (\cite{woo86}). Though $M_{{\rm Fe}}$ need
not be
monotonic in $M_{{\rm ZAMS}}$ (\cite{bar75}), there do seem to be
two
families of core structures at collapse: 1) low entropy $(<1)$, compact, low
mass iron or ONeMg cores with outer mantle density profiles that are steep and
2) high entropy $(\ge1)$, fatter, high mass cores with shallow density profiles
in their outer mantles. The latter generally have significantly more
intermediate mass elements (Si, Ca, Ar, S, etc.) at collapse and higher oxygen
yields. The oxygen yields increase systematically with ZAMS mass in a less
ambiguous way. Mass density profiles from \refmark{Weaver and Woosley
(1995)}{ww95}
representative of these two classes of cores are depicted in Figure 11. The two
families are clearly distinguished, though the specific ZAMS mass indexing
each profile should be given a lower significance. As we discussed in \S III,
the
smaller cores accumulate mass more slowly, but explode more easily (at least
in 1-D). The
relative ease with which the smaller cores might explode can be explained in
part by Figure 12, which depicts the behavior of the binding energy versus
interior mass. The total binding energies of the models are the intercept
values at $M=0$. However, the mass cut is further out (a neutron star or black
hole is left). For the fatter cores, the mass cut must be much further out for
a ``given'' amount of energy of order $1-2\times 10^{51}$ ergs, transferred
either hydrodynamically or by neutrinos, to yield an explosion. The larger
$\bfdot{M_s}$'s
associated with the larger cores may delay the explosion sufficiently so that
only lower neutrino luminosities are available to drive the explosion after it
commences (see \S V), resulting in an explosion energy that can not be much
higher than that for the smaller cores.
The delay to explosion for the more massive cores (if they explode) is
followed by a slower expansion as the larger envelope binding energies are
overcome, during which time the shocked matter can achieve
higher entropies via neutrino heating than can matter in an exploding core of
lower mass (\cite{wo94a}). These higher entropies might be relevant to the
r-process (\cite{mey92}) and may not be achievable in 1-D for the ``slimmer''
cores (cf. 15 $\mdot$)(\cite{burin}). The fat cores can explode only after
sufficient binding energy is accreted. If such cores reach the maximum mass of
a neutron star before exploding, a black hole is the likely result.
 At any rate, binding energy arguments
alone make it next to impossible for fat cores to yield the low-mass neutron
stars that are observed in binary pulsar systems (\cite{tho93};
\cite{arz94}). Furthermore, the fatter cores
generally explosively overproduce $^{56}$Ni by factors of 3--5, if the
SN1987A, SN1993J, and SN1994I data are representative (0.05--0.09$M_\odot$;
\cite{bou91}; \cite{nom93}; \cite{wo94b}). The overproduction of $^{56}$Ni is
correlated with the
overproduction in such models of intermediate mass elements (\cite{tim94}),
particularly of silicon. In addition, to avoid the
overproduction of neutron-rich species, the mass cut must be at or exterior to
the edge of the iron core, which for the fat models used here is at values of
1.74 and 1.78 $\mdot$, again too large to yield the measured pulsar masses
(but, for a possible way out, see \cite{ful94}).

It might be thought that the mass cut of the explosion is out in the freshly
synthesized $^{56}$Ni region and that much of the overproduction is buried in
the core (\cite{thn90}). However, such behavior might require
fine-tuning: the mass cut in ``realistic'' 1-D explosions is often determined
either at
explosion or as a result of reverse-shock reimplosion. To make matters
worse, the best fit oxygen yield of SN1987A is $\sim 1.3\ \mdot$
(\cite{chu94}; \cite{spy91}), just what some
of the models which overproduce intermediate mass elements and $^{56}$Ni say
the  6 $\mdot$ helium core of Sanduleak $-69^{\circ}$ 202 should produce
(\cite{woo88}; \cite{thn90}). One might be led to speculate that the fat cores
either can not
explode or if they explode must leave behind black holes or very fat neutron
stars
(\cite{bet94}). One may also speculate that
fat cores arise in stars with ZAMS masses higher than the $\sim$
18--20 $\mdot$ cut of models wwsb (Figure 10). The smaller cores of $18-20\
\mdot$ stars of
Nomoto and Hashimoto may be favored for SN1987A, but the pure Schwarzschild
condition that they employ in their calculations has recently been called into
question (\cite{sto94}). In a recent paper by
\refmark{Maeder (1992)}{mae92}, it is argued on the basis of the observed
${\Delta
Y\over\Delta Z}$ of the galaxy that there must be a maximum mass above which a
star can not be allowed to litter the ISM with all of its helium and
metals. This may be germane to the fate of the fat cores identified
above. Intriguingly, \refmark{Maeder (1992)}{mae92} suggests a cutoff ZAMS mass
of $\sim20-25\mdot$ (however, see \cite{pei94}).
(This conclusion depends sensitively on the IMF employed.)
In short,  there is as yet no consistent 1-D theory for massive
star evolution and explosion that fits all the data. The problems with the
progenitors are at least as severe as the problems with the 1-D explosion
mechanism itself. In particular, the radii and masses of the various $Y_e$,
entropy, and density ledges, that appear to be so sensitive to the convective
algorithm, deserve further scrutiny. Though the $M_{{\rm ZAMS}}$ to $M_{{\rm
Fe}}$
mapping is ambiguous, there do seem to be two classes of core structures with
which a supernova modeler can profitably contend and which may yield
different outcomes and products.

\section{The Neutrino-Driven Mechanism (One-Dimension)}

Fundamentally, the mechanism of a supernova explosion involves the transfer of
energy from the core that remains to the mantle that is ejected. In the prompt
mechanism, the piston action of the inner core was to do sufficient work on
the outer core to unbind it. The residual protoneutron star would have been
left more bound due to the work it performed. With the demise of the direct
mechanism, the agency of energy transfer is now thought to be neutrinos
(\cite{wil85}; \cite{bet85}). Core electron- and antielectron-type neutrinos
radiated from the neutrinospheres (see Figures 5a and 5b) deposit a
small fraction of their energy in the outer shocked mantle as they emerge. The
major
processes are $\overline\nu_e(p,n)e^+$ and $\nu_e(n,p)e^-$. The inverse
processes of electron capture are generally weaker near the shock and stronger
near the neutrinospheres. Neutrino-electron scattering can also heat the
matter. We describe in this section some of the basics of this process that
has yet to be definitively understood. In particular, without a convective
boost in the driving neutrino luminosities or other convective effects,
detailed 1-D calculations have failed to date to yield supernovae
(\cite{bur87};
\cite{may88}; \cite{bru87}; \cite{bru92}; \cite{bf93}; \cite{her92}). The
reader can refer to \refmark{Janka and M\"uller (1993)}{jm93}, \refmark{Janka
(1993)}{jan93}, \refmark{Burrows and Goshy (1993)}{bg93}, and \refmark{Bethe
and Wilson
(1985)}{bet85}  for a more rounded perspective of 1-D neutrino-driven
explosions.

The duration of the delay between the stall of the shock and its
revitalization is unknown and could range between 50 milliseconds and a
second. Delay has virtues: it allows low $Y_e$ material with $\eta$'s
$(=1-2Y_e)$ greater than 0.01 to be buried, it fattens the residue so that
observed neutron star gravitational masses $(M_{\rm G}\sim 1.35\ \mdot)$ can
obviously be
achieved, and it buries material with a high binding energy that would
otherwise have to be overcome. However, if the delay is too long, once the
supernova is launched the driving luminosities may have decayed too much to
yield supernova energies in the required range $(1-2\times 10^{51}$ergs). It
is yet to be determined whether a long delay is good (\cite{wo94a}) or bad
(\cite{tak94}) for the r-process,  which seems to require high entropies
(\cite{mey92}).  As stated in \S IV, in 1-D low-mass massive stars with thin
envelopes may not be able to achieve the requisite entropies.

As \refmark{Burrows and Goshy (1993)}{bg93} and \refmark{Burrows
(1987)}{bur87} have shown,
the supernova is a coronal phenomenon: when temperatures reach ``escape''
temperatures a wind is driven off of the protoneutron star. The neutrino
Eddington
luminosity is $\sim 10^{55}$ ergs/s, too large by one to two orders of
magnitude to be relevant. The protoneutron star bounded by an accretion shock
can (before explosion) be treated quasi-statically like any star; in
particular, it can be subjected to a stability analysis. Burrows and Goshy
(1993) showed that in 1-D there is a critical curve in the $L_\nu$ versus\
$\bfdot M$
plane,  above which the protoneutron star is unstable to a bifurcation to an
explosive solution. ($\bfdot{M}$ is the accretion rate at a radius of 200
kilometers, always
exterior to the shock before explosion and not conceptually equal to
$\bfdot{M_s}$.
However, it is numerically very
close to $\bfdot{M_s}$ during the quasi-static phase.)
In this analysis, the shock radius was treated as the
eigenvalue of the quasi-static problem. Above the critical curve, no solution
to
the steady-state can be found. The critical $\bfdot M$ is roughly proportional
to $(L_{\nu_e}/M_s)^{2.3}$ and $R_s$ in the steady-state is roughly
proportional to $L^2_{\nu_e}/\bfdot M$. The gain
radius (\cite{may85}) exterior to which heating $(H)$ predominates over
cooling $(C)$ is where $H=C$. This is almost exactly at the entropy peak,
since in the steady state, $u_1T{ds\over dr}\cong H-C$, where $u_1$ and $T$
are the post-shock settling velocity and the temperature, respectively, and
$s$ is the specific entropy. {\it Before} explosion, the integral (exterior to
the
electron neutrinosphere) of the cooling function is always {\it greater} than
that of the heating function. {\it Before explosion, the energy deposited in
the gain region by neutrinos bears no obvious relationship to the explosion
energy}. In spherical accretion, it is advected into the core and
reradiated. At the onset of instability (explosion), the ``optical'' depth to
$\nu_e$'s at the gain radius is 0.05 to 0.15, not 0.01, and the peak entropy
is less than 25 (\cite{bg93}). As the explosion
develops, the optical depth
decreases. {\it Before} explosion, oscillations of the shock (\`a la AM Her
stars (\cite{lan81}; \cite{che82})) are
overcritically damped both by neutrino losses and shock damping (with a
characteristic time that scales with $M_s/\bfdot M$) and the matter is
{\it never} radiation-dominated.

Though the eigenvalue analysis of \refmark{Burrows and Goshy (1993)}{bg93}
assumed
nothing about the actual neutrino luminosities, they were able to show that
accretion power alone could not ignite a supernova in spherical symmetry. The
critical $L_\nu$ vs $\bfdot M$ curve intersected the $L_{{\rm accretion}}$ vs
$\bfdot M$ curve at values of $\bfdot M$ that were higher by as much as an
order
of magnitude than those obtained after the stalled shock settles (see Figure
7). To produce a neutrino-driven explosion in ``spherical'' symmetry, some
other source of luminosity must be available, be it due to rapid diffusion out
of the core (modified neutrino opacities) or ``convective'' enhancement.

The rate at which heat is deposited
via the charged-current absorptions on free nucleons at a radius $r$ is
approximately given by,
$$\bfdot\epsilon_\nu=200{{\rm MeV}\over {\rm baryon}\cdot
s}\left({T_{\nu_e}\over 4.5
{\rm MeV}}\right)^2 \left({L_{\nu_e}\over 10^{52}{\rm ergs/s}}\right)
\left({10^2{\rm km}\over r}\right)^2,\eqno(5)$$
where it was assumed that $L_{\overline\nu_e}=L_{\nu_e}$, $Y_n+Y_p=1$ and
$T_{\overline\nu_e}=T_{\nu_e}$, the temperature of the electron neutrinosphere
(\cite{bet85}). At $L_{\nu_e}=4\times10^{52}$ergs/s,
$T_{\nu_e}=4.5$ MeV, and $r=10^2$ kilometers, the time to raise the
temperature of the matter by 1 MeV $(\tau\sim\epsilon_T/
\bfdot\epsilon_\nu)$ (including the electrons in the specific heat) is roughly
10 milliseconds. This is a measure of the time to achieve a steady-state or
the timescale of explosive expansion in $R_s$, once the mantle of the
protoneutron star is unstable. In the steady-state it is also ``equal'' to
${R_s\over u_1}$, where, again, $u_1$ is the post-shock infall speed. Since
$L_{\nu_e}\propto T_{\nu_e}^4R_\nu^2$, where $R_\nu$ is the neutrinosphere
radius,
$\bfdot\epsilon_\nu$ is proportional to $T^6_{\nu_e}({R_\nu\over r})^2$, a very
steep power of $T_{\nu_e}$. If we assumed that the cooling function is
proportional to $T^6$, the equilibrium matter temperature would be a weak
power of $r\ (\propto 1/r^{{1\over 3}})$. Actually, the advective term in
the heat equation forces the gradient in $T$ to be steeper in the steady state
(\cite{bru92}), but $T_{\nu_e}$ always provides an upper bound to the
shocked mantle temperatures and the gradient of the matter temperature
exterior to $R_\nu$ is always ``shallow.'' Since $T_{\nu_e}$ is $4-5$ MeV from
quite general neutrino opacity arguments, matter temperatures behind the shock
of $1-4$ MeV are easily understood. These are not the temperatures of 100 MeV
one would get by setting $\epsilon_g(={GM\over R})$ equal to $\epsilon_T$
(thermal) at a typical cold neutron star radius of 10 kilometers. The
``escape'' temperature, $T_{esc}$, that one gets by setting
$\epsilon_g=\epsilon_T$ is much larger than the real temperature near
the neutrinosphere. Since it decreases as $1/r$ and the matter temperature
actually decreases more slowly, the two curves intersect at some radius
exterior to $R_\nu$, typically 100--200 kilometers, at a value near 2--3
MeV. This temperature, not the core temperatures or $T_{\nu_e}$, sets the
specific energy scale of the explosion, and hence, along with the binding
energy of the progenitor envelope (Figure 12), sets the magnitude of the total
supernova explosion
energy. As with any corona, when the local
temperature reaches the ``escape'' temperature, a powerful wind is driven off
the star. Semi-quantitatively, the shock radius must be exterior to this
radius (at which $\epsilon_T\sim\epsilon_g$). These arguments can be shown to
be conceptually equivalent to those in the eigenvalue study of
\refmark{Burrows and Goshy (1993)}{bg93} when the entire star, with bounding
accretion shock, is analyzed.

When thinking about supernova explosion energies, a few basic facts are
useful. About a third of a solar mass of ideal gas nucleons at a temperature
of 1 MeV has an internal energy of $10^{51}$ ergs. A spherical volume
with a radius of 1000 kilometers filled with radiation
(with pairs) at a temperature of 1 MeV
has an internal energy of $1.6\times 10^{51}$ ergs. The thermonuclear energy
derived from the complete burning of $1\ \mdot$ of oxygen to $^{56}$Ni is
$10^{51}$ ergs and of $1\ \mdot$ of silicon to $^{56}$Ni is approximately
$0.4\times 10^{51}$ ergs. The burning of $0.07\ \mdot$ (87A!) of oxygen
completely to $^{56}$Ni yields only $0.7\times 10^{50}$ ergs.

Both \refmark{Janka (1993)}{jan93} and \refmark{Burrows and Goshy
(1993)}{bg93} have shown
that a large driving neutrino luminosity radiated over a short time (a flash)
is more efficient at igniting a supernova and ejecting the envelope than the
same total neutrino energy radiated over a longer time. It has been suggested
(\cite{bur87}) that such a flash might be the consequence of
convective motions near the neutrinosphere, but this has not been
convincingly demonstrated (\cite{bf93}). To simulate
an early
explosion with such a character, we have altered our model w15t by
artificially hardening the emergent electron neutrino spectra to have a
Fermi-Dirac $\eta$ of 3.0. Note that, guided by the
work of Myra and Burrows (1990), we assume in these calculations that the
emergent
$\nu_e$ and $\overline{\nu}_e$ spectra are Fermi-Dirac in shape, with a default
$\eta$ of
2 and a temperature equal to the matter temperature at decoupling (see \S II).
The velocity vs.\ radius
profiles of this model
(w15e3) are depicted in Figure 13. This figure shows the progress of the
supernova shock to a radius of $\sim$4000 kilometers and the collision of the
protoneutron star wind with the inner ejecta. The kinetic energy versus time
for model w15e3 is depicted in Figure 9. At the end of the calculation it has
reached $1.2\times 10^{51}$ ergs.
There are interesting differences between
models w15e3 and w15n, both of which exploded. In model w15e3, the peak
entropies achieved in most of the matter reach $\sim$34, while in model w15n
without neutrinos the number was 10--15 (Figure 14). In model w15n, when the
shock had
reached $\sim$4000 kilometers, most of the supernova energy was in kinetic
energy, while at the same radius in model w15e3 it was more evenly divided
between internal and kinetic energy. In neither blast was the internal energy
overwhelmingly in photons and pairs when 4000 kilometers was reached. The
post-shock temperatures in the two cases were interestingly different, with
consequences for explosive nucleosynthesis. In particular, the ansatz often
employed in explosive
nucleosynthetic studies that all the explosion energy is in radiation when the
silicon and oxygen zones are encountered will have to be reexamined
(\cite{auf91}; \cite{thn90}). A major
conclusion of the comparison of models w15e3 and w15n is that the mechanism of
explosion affects the subsequent nucleosynthesis, etc., even when the
explosion energies for the different explosion scenarios are comparable. Note
also that only a neutrino-driven explosion can yield entropies above 30.
If such entropies (and higher) are required by
nucleosynthetic arguments (\cite{mey92}; \cite{tak94}), then such arguments
are telling us something about the
supernova mechanism itself. Note that the entropy profile of model w15e3 at
late times is grossly Rayleigh-Taylor unstable. This means that the
calculations should have been done from an earlier time in more than one
dimension. This may be directly relevant to the high speed $^{56}$Ni bullets
observed in SN1987A (see \S VII). In addition, since (as stated in \S III) the
mass
between the shock and the neutrinosphere becomes small early after bounce
$(<0.01\mdot)$, explosion calculations in 1-D, such as w15e3, suggest that
the mass
cut might be determined at the {\it onset} of explosion, not later. The
outgoing
protoneutron star wind would assure this (modulo fallback due to reverse
shocks on much
longer timescales).

It was not our purpose in this section to explore the many interesting
consequences of 1-D models such as w15e3. These will be postponed to a later
paper. Rather, in this section we hoped to communicate a few of the salient
and important features of generic neutrino-driven explosions to establish the
context both of such a mechanism and of our new multi-dimensional calculations
discussed in \S VII.

\section{Instabilities and Convection in Supernova Explosions: Preliminaries}

Most hydrodynamic problems in astrophysics involve instabilities that are
either central or decorative, but are often ignored. Rayleigh-Taylor and
Kelvin-Helmholtz instabilities during supernova explosions are both of the
former and can no longer be the latter. The wildly varying composition,
entropy, density, and pressure gradients behind the supernova shock before and
after it explodes drive violent overturning in which the Mach numbers can
approach unity (\cite{bur87}; \cite{bl88}; \cite{bf92}; \cite{jm93a};
\refmark{Herant, Benz, and Colgate (1992, HBC)}{her92}; \refmark{Benz, Colgate,
and Herant 1994}{bch94};
\cite{fry94}; \cite{mfa91}). The question that is now being
addressed is whether such instabilities are crucial to the supernova mechanism
itself and attendant issues and on this there are contending opinions.

The basic physics of convection in supernovae was first seriously discussed by
\refmark{Epstein (1979)}{eps79}, who employed a mixing-length prescription and
emphasized lepton-driven convection. No conclusions concerning the role of
such motions were reached, but this work stimulated a flurry of activity in the
early 1980's (\cite{col80}; \cite{liv80}; \cite{bru79}; \cite{sma81}). In
these papers, lepton-driven, as opposed to entropy-driven,
instabilities were the focus, since it was known that the protoneutron star
would neutronize from the outside in (however, see \cite{sma81}). Negative
lepton gradients were supposed to drive either violent
overturn of the entire core or to accelerate the loss of lepton number and
energy. The paper by \refmark{Bruenn, Buchler, and Livio (1979)}{bru79} is
especially noteworthy
in
its suggestion that lepton-driven convection might enhance the neutrino
luminosity that could thereby ignite the supernova (see their model
5). However, though multi-dimensional hydro codes were used in some of these
papers, the initial models, transport, hydro algorithms, boundary conditions,
and physical assumptions often left much to be desired. With the more rigorous
work of \refmark{Lattimer and Mazurek (1981)}{lat81}, it was realized that
violent
core overturn was suppressed by both the shock-imposed positive entropy
gradients (Figures 3) and the shallowness of the lepton gradients in the inner
core (Figures 4), and interest in the role of hydrodynamic instabilities in
supernova explosions waned.  More recently, \refmark{Wilson (1985)}{wil85} and
\refmark{Mayle and Wilson (1988)}{may88} advanced a neutrino heating mechanism
(\S
V) that in their calculations required ``neutron finger'' instabilities to
sufficiently boost the driving neutrino luminosity. This ``neutron finger''
instability is {\it not} the lepton-driven instability, but is akin to the
salt-finger instability seen in the oceans (in particular at the mouth of the
Mediterranean) and relies for its existence on a large ratio of heat
diffusivity to composition diffusivity. In the terrestrial example, the
layering of hot, salty water over cold, fresh water, which is stable by the
normal Ledoux or Brunt-V\"ais\"al\"a analysis, is unstable because a dimple at
the slab interfaces allows enhanced heat transfer without a correspondingly
rapid transfer of salt. The result is ``cold'' salty water over cold fresh
water (a density inversion) and overturn, but its rate is slaved to the
relative rate of heat and salt diffusion. This is {\it not} the classic
convection, and laboratory experiments and ocean observations frequently show
that a layered structure, not churning mixing, results (\cite{tur74}). In the
protoneutron star case, it was thought that since heat was
transported by six neutrino species, whereas lepton number (``salt'') was
transported only by $\nu_e (+)$ and $\overline\nu_e(-)$, the hot neutron star
was similarly unstable. The low opacities of the $\nu_\mu$-matter interaction
were to enhance the effect, which in the calculations of \refmark{Mayle and
Wilson (1988)}{may88}  was
simulated with a mixing-length algorithm. That the matter
was importantly unstable to neutron finger instabilities was immediately
challenged by \refmark{Burrows (1987)}{bur87} and \refmark{Burrows and
Lattimer (1988)}{bl88} and has recently been challenged in
a more comprehensive work by
\refmark{Bruenn, Mezzacappa, and Dineva (1994)}{bmd94}. In the latter, it is
pointed out that once the ``$\nu_\mu$'s'' start to move quickly, they
uselessly decouple from the matter.

\refmark{Burrows (1987)}{bur87} suggested that the negative entropy gradient
originally created by the stalling shock would, because of its proximity to
the neutrinospheres, drive convective overturn that because of this proximity
would churn up heat, boost the neutrino luminosities, and by the neutrino
heating mechanism ignite the supernova.  This effect was shown to
exist (\cite{bur87}; \cite{bl88};
\cite{bf93}), but it now seems that it exists for
only about 10--20 milliseconds, since the transport of heat by neutrinos
smooths the gradients out (see Figures 3, \cite{bf93}, and
\cite{bru94}). However, as was shown by
\refmark{Burrows and Fryxell (1993)}{bf93}, right under and around the
neutrinospheres the lepton gradients in particular, but also the entropy
gradients, are unstable to standard convection, which should persist. Using a
mixing-length approach, Bruenn and Mezzacappa (1994) have suggested that this
effect
on the neutrino luminosities is marginal, but \refmark{Burrows and Fryxell
(1993)}{bf93} find using a 2-D
hydrodynamics code that the effect can be
important. When the Mach numbers and scales of the flow are so large, a
mixing-length calculation will underestimate such effects. In addition, the
work done by the hydrodynamic overturn can be large and can readjust the flow
(e.g. move the shock outward (\cite{bl88},
\cite{bf92})). These effects can not be adequately
treated by mixing-length.

\refmark{Bethe (1990)}{bet90} argued that the region just behind the shock
should  be
a theorist's focus and that neutrino heating ``from below'' of this material
would naturally create negative and unstable entropy gradients, whatever the
initial gradients. The material exterior to the gain radius would overturn and
the energy deposited by the core neutrinos, instead of being reradiated and
lost as the matter plummets into the pit, would be available to do work on the
shock. The stalled shock might thereby be revitalized and {\it driven} during
the explosion. \refmark{HBC}{her92} and
\refmark{Herant, Benz, Hix, Fryer, and Colgate (1994, HBHFC)}{hbh94} have
recently lent
support to this view in an extensive and important series of detailed 2-D
simulations. In a careful 2-D parameter study with spherical light-bulb
heating,
\refmark{Janka and M\"uller (1993a)}{jm93a} partially support
Bethe's general idea, but, along with \refmark{Burrows and Fryxell
(1993)}{bf93}
challenge some of HBC's notions of the flow structure. This outer
entropy-driven instability near the shock joins the inner entropy- and
lepton-driven instabilities and the neutron-finger instability as the major
instabilities suggested to be important in reviving and energizing a stalled
shock.  The effects of rotation have been addressed by \refmark{M\"onchmeyer
and M\"uller (1989)}{mon89}, \refmark{Shimizu \etal\ (1993)}{shi93}, and
\refmark{Shimizu \etal\ (1994)}{shi94}.  The latter emphasize
the rotation-induced asymmetry of the neutrinosphere, which the authors
contend will produce jets and large pulsar kick velocities. \refmark{M\"uller
(1993)}{mul93} has
compared 2-D flow structures with the 3-D flow structures and finds
that overshooting is weaker in 3-D. It is important to understand the
differences between 2-D and 3-D flow behavior, but, currently, full 3-D
rad/hydro simulations are beyond the reach and patience of supernova modelers.

Before we turn in \S VII to the results of our new two-dimensional supernova
simulations, a few simple analytic arguments can help guide the discussion.
The Brunt-V\"ais\"ala angular frequency $N$ is given by
$$N^2=-g\left({\partial ln \rho\over\partial r}-{1\over \gamma}{\partial ln
P\over \partial r}\right),\eqno(6)$$
where $g$ is the effective gravity defined to be positive in the negative $r$
direction and $\left.\gamma={\partial lnP\over \partial ln
\rho}\right|_s$. This is
exactly the frequency derived in the linear, compressible Rayleigh-Taylor
analysis. The two are equivalent and the standard incompressible
Rayleigh-Taylor result is seen in the $\gamma\to\infty$ limit. Eq.\ (6) can be
rewritten using a few thermodynamic identities (\cite{lat81}) to yield,
$$N^2=\left.{g\over\gamma}\left[{\partial lnP\over \partial
s}\right|_{\rho,Y_e}
{ds\over dr}+\left.{\partial ln P\over \partial Y_e}\right|_{\rho,s}
{dY_e\over dr}\right],\eqno(7)$$
where, for specificity, we have ignored the effects of neutrinos. The
neutrinos can easily be included, but care must be taken when moving from the
neutrino-opaque, beta-equilibrium region to the neutrino-transparent
region. Eq.\ (7) shows that when (as is common) the pressure derivatives are
positive, negative entropy and $Y_e$ gradients are unstable
$(N^2<0)$. $\left.{\partial lnP\over \partial Y_e}\right|_{\rho,s}$ can be
negative, but only at sufficiently high $s$'s and low $Y_e$'s to be rare
(\cite{lat81}). Note that $\left|{\partial lnP\over
\partial s}\right|_{\rho,Y_e}={1\over C_v}$, where $C_v$ is the specific heat
at constant volume.
The overturn timescales from this linear analysis in the protoneutron star
context are expected to be
3--10 milliseconds. Eq.\ (7) encapsulates the Ledoux criterion for convection.

Hydrostatic equilibrium structures are minimum total energy structures,
subject to the constraint of mass conservation. In a more general
thermodynamic sense, they are minimum {\it free energy} structures. The
constraint of sphericity may inhibit the object's attempt to achieve this
minimum free energy. When $N^2$ in eq.\ (7) is negative, the structure can
achieve a lower free energy via convective overturn. The free energy
difference between the relaxed 1-D (spherical)-configurations and
``3-D''-configurations is available to do useful work. This work can help push
the shock outward, {\it even if a concommitant neutrino luminosity boost is
ignored}. Thus,
all convective motions are reservoirs, temporary or otherwise, of useful
energy.
This can be seen for the region of negative entropy created by
the stalling shock in the calculations of \refmark{Burrows and Fryxell
(1992)}{bf92}. There it is shown that, due only to overturn, the shock can
receive a significant boost.

If an unstable region is modeled as two constant-density slabs of equal
thickness, $\Delta x$, then it can easily be shown that the gravitational
energy available due to the simple interchange of the slabs is,
$$
W={\rho_1-\rho_2\over \rho_1+\rho_2}g\Delta M\Delta x,\eqno(8)$$
where $\Delta M$ is the total mass of the unstable region of thickness
$2\Delta x(=\Delta h)$ and $g$ is the local gravitational acceleration. If we
assume that $\rho_1-\rho_2={d\rho\over dr} {\Delta h\over 2}$ and that
$\rho_1$ and $\rho_2$ are not very different, we obtain
$$W\sim{1\over 8} {d\ ln\rho\over dr} g\Delta M (\Delta
h)^2.\eqno(9)$$
Using the correspondence between eqs.\ (6) and (7), we can substitute ${1\over
\gamma C_v} {ds\over dr}$ for ${dln\rho\over dr}$ in eq.\
(9) to find,
$$W\sim{1\over 8}{g\over\gamma C_v} {ds\over dr} \Delta M (\Delta
h)^2,\eqno(10)$$
where, for simplicity, we have ignored the ${dY_e\over dr}$ term. From eq.\
(10), we can derive two interesting relationships. The first gives simply and
very approximately the energy available in the overturn of a region of mass
$\Delta M$ and thickness $\Delta h$ with a given unstable $\Delta s$ at a
distance $r$ from the center of a protoneutron star of mass $M$:
$$W\sim 10^{51}{\rm\ ergs} \left({\Delta M\over 0.1\mdot}\right)
\left({\Delta h\over 10^2{\rm\ km}}\right) \left({\Delta s\over 10}\right)
\left({135{\rm\ km}\over r}\right)^2 \left({M\over 1.3\mdot}\right).\eqno(11)$$
We see that if $\Delta M\sim 0.3\mdot$, and $\Delta s\sim 5$, $W$ is of
order $10^{51}$ ergs. These numbers are realistic for the region behind the
stalled shock with the initially unstable entropy gradient.

Note again that this is free energy available, ignoring neutrino heating. If on
the other hand, the unstable $\Delta s$ is created by neutrino heating (as in
Bethe convection), eq.\ (10) can be transformed, assuming that $\langle
T\rangle \Delta s={\Delta E_\nu\over\Delta M_2}$ (where $\langle T\rangle$ is
an average temperature in a heated region of mass $\Delta M_2$ and $\Delta
E_\nu$ is the neutrino heat deposited), into
\begin{eqnarray*}
W &=& {g\over 8\gamma} {\Delta s\over C_v} \Delta M\Delta
h={g\over 8\gamma} {\Delta E_\nu\over \langle T\rangle \Delta M_2} {\Delta
M\Delta h\over C_v}\\
&=& {\Delta E_\nu\over \langle T\rangle} {\Delta M\over \Delta M_2} \left[
{g\Delta h\over 8\gamma C_v}\right].
\end{eqnarray*}
If we use the Bernouilli integral $({1\over 2} v^2_i+\Phi (r_i)+ {\gamma
P_i\over (\gamma-1)\rho_i}={\rm const.})$ for matter flowing from the shock to
the heated region (where it {\it presumably} turns around) and assume that
$v_1=v_2$, that $g\Delta h=\Delta\Phi$, that $P=\rho RT$, that $C_v={3\over
2}R$, that $\Delta M_2={1\over 2}\Delta M$, and that $\langle
T\rangle={T_h\over 2}$, we obtain
$$W={1\over 2}{\Delta T\over T_h}\Delta E_\nu,\eqno(11)$$
where $T_h$ is the temperature at the gain radius and $\Delta T$ is the
temperature difference between the shock ``slab'' and the gain radius
``slab.'' Eqs.\ (8) and (11) were derived by assuming the interchange of slabs
of thickness $\Delta h/2$. If instead, we say that the cycle is from the top
of the top slab (at the shock) to the bottom of the bottom slab (at the gain
radius), we obtain the Carnot relation,
$$W={\Delta T\over T_h}\Delta E_\nu.\eqno(12)$$
Notice that the $\gamma$ has dropped out. A relation such as eq.\ (12)
might suggest a more general validity. {\it Very crudely}, eq.\ (12) would
give the optimum
efficiency with which the deposited neutrino energy can be converted into
useful work. Such a relation was implictly posited in \refmark{HBHFC}{hbh94}.
However, this relation assumes that heating occurs on the ``power stroke'' and
not at all during early compression and that matter turns around at the gain
radius. We find in our 2-D simulations that neither of these assumptions is
particularly good. Furthermore, the analysis above ignores electron capture
and composition changes which turn out to be important. The
``convective engine'' is very leaky. In addition, the implication
of the engine analogy that the work done in a cycle has something to do with
the explosion condition
and the explosion energy is misleading.  Since the boiling phase before
explosion is a succession of
quasi-steady states, neither the total neutrino energy deposited nor the
work done in putative cycles during this waiting phase are remembered by the
matter or are
relevant to the subsequent explosive evolution.

\section{New Multi-Dimensional Supernova Explosion Results}
To focus the discussion and to avoid juggling an avalanche of plots and
numbers, we highlight in this paper only one full 2-D simulation of a model of
a massive star progenitor, s15s7b of Weaver and Woosley (1995). This 15 $\mdot$
model was discussed previously in the context of models w15t and w15n and is
typical of the compact class with smaller iron cores $(M_{Fe}\sim 1.28\ \mdot)$
lighter envelopes (Figure 11), and lower binding energies (Figure 12). With
our default neutrino transport scheme, this core does not explode in 1-D (see
Figure 2a). However, as we will demonstrate, in 2-D the character of the flow
and the final outcome are qualitatively different.

We refer to the two-dimensional simulation of the full radiation/hydrodynamic
evolution of the core of the s15s7b model as our {\bf star}
run. The inner 4500 kilometers was included on the grid, which was 500
(radial) times 100 (angular) zones. Spherical coordinates with azimuthal
symmetry were used, and the angular zones were distributed evenly in the
$45^{\circ}<\theta<135^{\circ}$  interval (see, however, Janka and M\"uller
1993a).
The inner 15 kilometers was followed in
one-dimension using our restricted-2-D algorithm. The angular boundary
conditions were periodic. To resolve the pre-explosion convective stage, we
put 300 radial zones interior to 150 kilometers and distributed the remaining
200 radial zones in a quadratically expanding fashion so that the outer zone
was $\sim$50 kilometers thick $({\delta r\over r}\sim 0.01)$.

The {\bf star} run can be compared to the 1-D w15t run and its outcome. In
model w15t, the core bounced at $\sim$209 milliseconds (Table 2) and the shock
quickly settled to a radius of $\sim$80 kilometers (Figure 5a). (Note that,
unless otherwise
stated, the quoted times are the times since the start of the calculation, not
the times
since bounce. This is the common, if unfortunate,  convention.) Electron
capture in the shocked zone was copious and low $Y_e$'s  were achieved in a
trough between 0.8 and 1.3 $\mdot$ (Figure 4a). Within 60 milliseconds of
bounce, the shock radius was at $\sim$ 60 kilometers. The small shock radius
almost guarantees failure since the gain region is either
small or non-existent, electron capture is faster at the consequently higher
post-shock densities, and the accretion ram pressure is larger, for a given
$\bfdot M$. In model w15t, the shock radius continues to sink and there is no
explosion. A black hole would eventually form.

Since a
finite-difference code such as PPM is almost noiseless, we have to add seed
perturbations to start any
hydrodynamic instabilities. For the 2-D {\bf star} calculation, this
perturbation
was in density and was sinusoidal with an amplitude of 2\% and a period
of $\pi/5$. An additional random perturation of 1\% was superposed.  Memory
of the specific perturbation was completely lost after $\sim$10 milliseconds.
In sprinkling the seed perturbations, we were
guided by the convective oxygen burning study of \refmark{Bazan and Arnett
(1994)}{baz94}.  In that paper, density fluctuations as large as a few percent
and Mach numbers near 0.25 were obtained. We assumed that similar
excursions from quiescent mixing-length parameters obtained during the silicon
burning phase that immediately precedes core collapse. Since the inner edge of
the convective silicon burning shell is always between 0.8 and 1.1 $\mdot$
(interior) in the progenitor models and these zones are always encountered
within a millisecond of bounce, we get an immediate
start to the Rayleigh-Taylor instabilities that is physically motivated.

It is difficult to communicate all the characteristics and behaviors of a 2-D
hydrodynamic simulation in the standard paper format. We have derived
velocity, $Y_e$, temperature, entropy, pressure, Mach-number, composition, and
density (etc.) maps and profiles and their evolutions, but we can not hope to
convey all the nuances of the flow in only words or tables. Therefore, we
present in this paper representative color snapshots from the {\bf star} run of
contour plots of various quantities at interesting and illustrative phases of
the simulation, both before and after explosion.

Table 3 lists the 2-D plots that we have included to illustrate our
discussion.
\medskip
\centerline{\bf a. Multi-Dimensional Hydrodynamics Before Explosion}
\smallskip
Figure 15 depicts the inner 150 kilometers of the distribution at $t=238$
milliseconds (about 30 milliseconds after bounce). Velocity vectors that trace
the flow are superposed. The mildly aspherical shock is at the juncture of the
blue $(Y_e\sim 0.25-0.35)$ and red $(Y_e\sim 0.48)$ regions at radii between
84 and 94 kilometers. Due to electron capture on the shock-liberated protons,
$Y_e$ plummets to near 0.1 at $\sim$30 kilometers in a fashion similar to
that depicted in Figure 4a for model w15t. The immediate post-shock
temperatures and densities are $\sim$2 MeV and $\sim 10^{10}$ gm/cm$^3$,
respectively. The pre-shock temperature is near 0.7 MeV. As Figure 15 shows,
the material between the shock and $\sim$ 45 kilometers is unstable, as is the
region between 15 and 30 kilometers. Convection in the latter region is driven
by negative lepton and entropy gradients as discussed in \S VI, but is weak in
the {\bf star} series. The spherically distributed accretion flow exterior to
the shock is channelled into predominantly downward-moving plumes that hit the
inner core near the electron neutrinosphere in a very aspherical fashion. The
mass accretion flux into the inner core can vary with angle by factors of three
and the accretion neutrino luminosity varies accordingly. There are hot spots
of neutrino emission that dance over the inner core. Before the explosion, the
mass accretion rate through the shock is almost equal to that on the inner
core. As a consequence, the local mass flux per unit area at the
neutrinospheres and
the local neutrino fluxes can be much larger than in the 1-D calculations.
Figure 16a depicts the variation of the neutrino
luminosities (fluxes times area at large $r$) with angle at various
times. Figure 16b portrays the variation of the angle-averaged neutrino
luminosities versus time for both the {\bf star} calculation and its 1-D
analog. The
flickering depicted in Figures 16 is significant  and
dwarfs in magnitude the convective boost championed by \refmark{Burrows and
Fryxell (1993)}{bf93}.

In the {\bf star} calculation, the pre-explosion, post-bounce phase lasts
almost
100 milliseconds ($\sim$ 30 turnover times). During this time, the character of
the unstable flow
changes. Figures 17 through 21 depict this evolution in $Y_e$ maps at $t=270$,
299, 304, 307, and 311 milliseconds. The initial instability is a result of
the negative entropy gradient behind the shock, imposed as it stalls
(\cite{bur87}). As the neutrinospheres recede, the unstable region grows until
the region between the shock and $\sim$ 35 kilometers is engulfed. This
thickening of the unstable region can be seen by comparing Figures 15 and 18
at 238 and 299 milliseconds, respectively. As suggested by \refmark{Bethe
(1990)}{bet90} and \refmark{HBC}{her92}, after about 10 milliseconds into the
instability, it is neutrino heating from below that drives the subsequent
``convective'' motions. The flow takes on the character of vigorous
``boiling,''  but before explosion there is always a net mass flux through the
region. There is no secular tendency for the shock to move monotonically
outward during
this pre-explosion phase.
The convective region encompasses 4--6 pressure scale-heights.

Early after bounce, the flows interior to the shock are predominantly inward,
at speeds of $1-2\times 10^4$ km/s. There are some partially stagnated
accretion streams at lower speeds and a few upward plumes and swirls at a few
times $10^3$ km/s. As time advances towards explosion, more vigorous
convective eddies appear and along with the downward plumes, upward bubbles
rise
with speeds that may exceed $3\times 10^4$ km/s. Their Mach
numbers are near one. However, despite the bubbles, at all times before
explosion the average energy
and mass fluxes are {\it inward}.
As the external accretion rate decays, the unstable
motions become more vigorous. Early in the convective phase, the bubbles
created by neutrino heating are at lower entropies of 10--15, but those formed
later on can reach entropies of 25--35. These local entropies are larger than
can be achieved before explosion in 1-D (\cite{bg93}). In fact, small hot blobs
with entropies near 60 are formed when the rising bubbles encounter the shock
and are turned around. Dwelling in the gain region longer than is possible in
1-D allows small clouds of gas to achieve entropies unachievable in 1-D before
explosion. However, the swirling motions return these hot blobs to the cooling
regions where their entropies drop below 35.

A major feature of the unstable flow is the formation of many bubbles that
rise and collide with the shock. These collisions warp and bow out the shock
surface episodically on timescales of 1--5 milliseconds. The shock radius can
vary with angle by as much as 30\% (see Figure 18). The bubbles give the shock
a botryoidal appearance, much like an aggregate of hematite or an oscillating
drop of charged liquid mercury. In addition, the average radius of the shock
varies between 85 and 120 kilometers in response to the rising bubbles and the
gradual decay of the mass accretion rate. This is in contrast to a 1-D shock
radius below 60 kilometers and is an index of the important difference between
one- and
multi-dimensional simulations. This difference is most clearly seen in Figure
5c in which the 1-D and 2-D shock radii versus time are depicted.
In 2-D, a gain region of respectable size is
maintained so that when the accretion rate has subsided sufficiently, the core
can explode into a supernova, as the {\bf star} series eventually does near 310
milliseconds (Figures 20 and 21).

When a particularly vigorous bubble encounters the shock, if the explosion is
not imminent, the bubble material is forced along the shock and then
downward and can encounter similar streams of matter to form a coherent
columnar downflow.
(During this collision, the shock is pushed to slightly larger radii.)
Such plumes are seen in Figures 19--21 and similar plumes have been identified
by HBC. The entropy of matter in these downward plumes is generally ten to
fifteen units lower than in the bubbles. Figure 22 depicts the entropy
structure at $t=304$ milliseconds when such a plume is in evidence. The plume
is as much as four times denser than adjacent matter at the same
radius. However, plumes seen earlier in the calculation broke up within 5--10
milliseconds and the flow resumed its more chaotic appearance.
We do not see a final merger of small eddies into larger ones,
or the appearance with time of a dominant ``$\ell=1$'' mode. The flow is
always fluctuating and complicated,
with equivalent $\ell$ mode numbers of 2--10 generally in evidence. We never
see families of simple convective cells. The origin of the differences
between
our results and those in the \refmark{HBC}{her92} and HBHFC series is not clear
to us.
(The two groups do almost everything differently, with different base hydro
codes). However, the fact that we allow the neutrino fluxes to vary with angle
is certainly a factor, as is the fact that our shock stalls at smaller radius.

Difficult to see in Figures 15 through 21, but obvious in our analysis, are
quasi-periodic {\bf gravity waves} (g-modes) at the base of the unstable
region near the neutrinospheres. These internal modes propagate
perpendicularly in the theta direction with
speeds of five to fifteen kilometers per millisecond and with a variety of
periods
around 3 milliseconds. They have wavelengths between 5 and 20
kilometers. Their displacement amplitudes can be greater than 5 kilometers
and as the gravity waves propagate over the neutrinospheres they ripple them
perceptibly (Figure 23). Penetration of the neutrinosphere by as much as 5--10
kilometers is a result. The existence and persistence of numerous strong
gravity modes is another major hydrodynamic feature of this pre-explosive
phase and is correlated with the angular variations in the neutrino fluxes and
the undulating motions of the accretion shock. In fact, there seems to be a
strong coupling between the meandering columnar plumes and the g-modes that
they excite. Such feedback has been seen in other simulations of compressible
convection (\cite{hur86}). The boost in the average
neutrino luminosities due to the overshoot and dredge-up is difficult to
distinguish from the wildly varying fluxes due to aspherical accretion, but
from a comparison with the 1-D calculation seems to grow to be as much as 20\%
just before explosion (Figure 16b).

Before explosion, we see no evidence of an ``accumulation'' or ``building'' of
total energy.  In fact, in the {\bf star} calculation before explosion, but
after bounce,
the total energy
(kinetic plus internal plus gravitational) in the overturning region
actually decreases
(though the {\bf star} model does eventually explode!).
Since $\bfdot{M}$ changes only slowly,
the ``boiling'' shocked material is in quasi-hydrostatic equilibrium.
Furthermore,
the average shock radius does not increase monotonically with time (see Figure
5c).
If $\bfdot M$ is held constant, the average shock radius and the kinetic and
thermal energies are roughly constant and there is no tendency to
explode. We checked this by putting a constant density, velocity, pressure, and
energy
outer boundary condition at 200 kilometers about 50 milliseconds after bounce.
The mass accretion rate stabilized and the increase in the kinetic energy
stopped.  There was no explosion.  The increase in ``convective'' vigor
with time that we see in the simulations is in some sense in
response to the decay of $\bfdot M$ and the accretion ram.

It is important to note that the accuracy of the neutrino transport
algorithm employed can make a qualitiative difference in the outcome. Our
``st2d'' series used different approximations for the neutrino-matter coupling
above the neutrinospheres and the resulting accretion shock radii were as much
as ten kilometers smaller. This ``st2d'' series did not explode. Another
series that we performed (the ``bang'' series) employed a neutrino transfer
algorithm that enhanced the neutrino-matter coupling in the semi-transparent
region. Its pre-explosion accretion shock radii varied between 100 and 150
kilometers, it exploded within 50 milliseconds, and it involved less vigorous
convective motions, that did not plunge all the way to the neutrinospheres. In
the ``bang'' simulation, two uncoupled convective regions emerged, an inner one
driven by shallowly negative entropy and lepton gradients between 20 and 50
kilometers and an outer one driven \`a la Bethe by neutrino heating between
100 and 150 kilometers. The inner convection boosted the core neutrino
luminosity by as much as 10\% (\cite{bf93}), but does not seem to have been the
crucial factor in igniting the explosion. (This model almost exploded in 1-D!)
The lesson we draw from these ancillary studies is that, though ``Bethe''
convection helps a core to explode, it does not guarantee it. Bethe convection
results in larger accretion shock radii and gain regions (see \S VI), but
neutrino energy and lepton losses and the accretion ram can still be too much
to overcome. Explosion is a quantitative question that hinges for the theorist
on the physical assumptions employed. In particular, a model with the best
neutrino transport, but a dense envelope and large consequent $\bfdot M$'s need
not explode, even when multi-dimensional effects are included.
We find that Bethe convection exists and is very useful,
but that it does not guarantee explosion.
\vskip8truept
\centerline{{\bf b. The Explosion}}
\smallskip
As we have stated, we believe that whether a ``theoretical'' core explodes is
a quantitative question. Our default 2-D calculation of s15s7b, the
{\bf star} simulation, exploded magnificently 100 milliseconds after
bounce.
We are compelled to ask why. Figure 20 shows the
flow field at $t=307$ milliseconds, a few milliseconds before explosion. The
shock is between 100 and 120 kilometers in radius, and there are two strong
downward plumes near 5$^{\circ}$ and 45$^{\circ}$. The ``boiling'' is more
vigorous than
50--100 milliseconds earlier and the excursions in ${\Delta R_s\over R_s}$ are
a bit larger. However, in many ways, nothing extraordinary is occurring. Yet,
by 311 milliseconds (Figure 21), the core is obviously exploding. Just before
the explosion, the shock moves inward in places by about 10 kilometers and the
accretion luminosity rises fractionally, but it does so undramatically. We are
led to conclude that {\it just as in the 1-D case} described in \S V the
shocked
envelope reaches a {\it critical state} and becomes unstable.  Note, however,
that the 1-D calculation w15t did not explode, despite
the steady decay of the mass accretion rate. ``Convection'' is obviously an
important ingredient in this supernova explosion. Before the explosion, the
positive influences of neutrino heating and convection are ``balanced'' by the
negative effects of electron capture, neutrino energy losses, and the
accretion ram. Eventually, when the critical condition is reached this balance
is tipped in favor of dynamical
expansion.  Note that, at the onset of explosion, the matter that will
eventually be
ejected may still be {\it bound}.  Neutrino heating during the explosion is
crucial
to the eventual achievement of supernova energies.  The aggregate energy
deposited
before explosion is almost completely irrelevant to the final supernova energy.

Since the precise nature of the trigger is so important, we digress a bit here
to
summarize the role of 2-D in triggering the explosion.
In 1-D, the heated
parcel would perforce fall directly into the cooling region interior to
the gain radius (where cooling = heating) and lose its just recently
gained energy. (Curiously, the cooling region lies closer to the
neutrinospheres where the temperatures are higher.)
However, the rising balloons, upon
encountering the shock, are immediately advected inward by the powerful mass
accretion flux raining down. They do not dwell near the shock. In fact, the
net mass flux through the shock is approximately equal to the mass flux onto
the core and mass does not accumulate in the convective zone. The mass between
the shock and the neutrinospheres {\it decreased} by about a factor of three
in the {\bf star} calculation during the pre-explosion boiling
phase
that lasted $\sim$100 milliseconds ($\sim$30 convective turnover times).
All the matter that participates in the ``convection'' before explosion
eventually leaves the convection zone and settles onto the core. A given
parcel of matter may ``cycle'' one or two times before settling inward (and a
large fraction never rises), but more than three times is rare. The boiling
zone is resupplied with mass by mass accretion through the shock and a
secularly evolving steady-state is reached. This steady-state is similar to
that achieved in 1-D, but due to the higher dwell time the average entropy in
the envelope is larger and its entropy gradient is flatter. These effects,
together with the dynamical pressure of the buoyant plumes, serve to increase
the steady-state shock radius over its value in 1-D by 30\%--100\% (Figure 5c).
It is this
effect of boiling that is central to its role in triggering the explosion, for
it thereby lowers the critical luminosity threshold (Burrows and Goshy 1993).
The
lowering of the effective critical curve allows the actual model trajectory in
$L_{\nu_e}$ vs. $\bfdot M$ space to intersect it. Even if in 1-D it can be
shown
that the two curves can intersect, they would intersect earlier and more
assuredly with the multi-dimensional effects included. The physical reasons
for the lowered threshold are straightforward: a large $R_s$ enlarges the
volume of the gain region, puts shocked matter lower in the gravitational
potential well, and lowers the accretion ram pressure at the shock for a given
$\bfdot M$. Since the ``escape'' temperature $(T_{\rm esc}\propto{GM\mu\over
kR})$
decreases with radius faster than the actual matter temperature $(T)$ behind
the shock, a larger $R_s$ puts a larger fraction of the shocked material above
its local escape temperature. $T>T_{\rm esc}$ is the condition for
a thermally-driven corona to lift off of a star. In one, two, or three
dimensions, since supernovae are driven by neutrino heating, they are coronal
phenomena, akin to winds, though initially bounded by an accretion
tamp. Neutrino radiation pressure is unimportant.

We conclude that the instability that leads to explosion in $\ge$2-D is of the
same character as that which leads to explosion in 1-D. Since the explosion
succeeds a quasi-steady-state phase, neither the total neutrino energy
deposited during the boiling phase nor any putative coeval thermodynamic cycle
is of relevance to the energy of the explosion or the trigger
criterion. Energy does not accumulate in the overturning region before
explosion (it in fact decreases) and the increasing vigor (speed) of
convection is in response to the
decay of $\bfdot M$. If $\bfdot M$ were held constant, the overturning would
not
grow more vigorous with every ``cycle'' and a simple, stable convective zone
would be established. In fact, before explosion the average total energy
fluxes $((\epsilon+P/\rho+{1\over 2}v^2-{GM\over r})\bfdot{M})$ due to the
overturning motions are {\it inward}, not outward, since the net direction of
the matter is onto the core.

The average electron-neutrino luminosity of the {\bf star} model at explosion
was lower than that required by the 1-D theory of \refmark{Burrows and Goshy
(1993)}{bg93}. {\it However, a major consequence of the breaking of spherical
symmetry and the
resulting increase in the steady-state shock radius
is the lowering of the critical explosion threshold}. In addition, as we have
shown, the fluctuations in the neutrino
fluxes are quite large (Figures 16). Since in 1-D it has been demonstrated that
a ``flash'' of neutrinos is more efficient at igniting an explosion than the
same energy radiated over a longer time (\cite{jan93}), it could be that the
observed variation in $L_\nu$ with angle plays a role in triggering
the explosion.
In addition, as Figure 16b implies, the effects of dredge-up on the average
luminosities are not small just before the explosion. Once the explosion
commences, it runs away, since
neutrino cooling is a stiff function of temperature and the temperature
decreases  slightly as the
envelope expands. Though the analytic development that led to equations 8--12
is of qualitative interest, the flow complexity demonstrated in Figures
17--21, the quasi-steady-state nature of the pre-explosion object,
the influence of the g-modes, copious electron capture, the
simultaneous existence of bubbles and downward accretion plumes, and the
heterogeneous entropy structure in the pre-explosion envelope make its serious
application of dubious use. The ``convective engine'' analogy
\refmark{(HBHFC)}{hbh94} does not
capture the essential physics of the phenomena we see.

Figures 24 through 30 depict the development of the explosion from 318 to 408
milliseconds. Figures 24 through 27 show the entropy distribution at 318, 348,
378, and 408 milliseconds, while Figures 28, 29, and 30 show the $Y_e$, radial
velocity, and log $\rho$ distributions, respectively, at $t=408$
milliseconds. Also displayed on Figure 28 are velocity vectors that trace the
flow. This set of figures encapsulates a rich variety of results and
conclusions.

When the explosion commences, the high entropy bubbles drive it and
they are not distributed isotropically. Plumes and {\bf fingers} (some at times
almost jet-like) emerge from the core and push the shock outward. The working
surfaces of the plumes near the shock spread out and back in the classical
fashion of jets and experience Kelvin-Helmholtz instabilities not well
resolved in these calculations. The anisotropy of the material behind the shock
is much larger than that of the shock itself, which may be only 10--20\%
$(={\Delta R_s\over R_s})$. The entropies of the fingers vary from 10 to 40,
even at the same radius. The vigorous finger at $\sim60^{\circ}$ has
entropies that range from 20 to 40, Mach numbers near 3, outward velocities
as high as 50,000 km/s
(Figure 29), and densities that can be a factor of four smaller than those of
adjacent matter. The shock wave itself moves outward at speeds from 20,000 to
25,000 km/s.
The density distribution is very heterogeneous, with low and
high density regions at the same radii and regions in which the density at
small radii is smaller than the density at larger radii, along the same angular
ray. In our 90$^{\circ}$ wedge, at least three fingers are in evidence.
Such fingers may be part of the explanation for the high
velocities and asymmetrical gamma-ray and infrared line profiles of the
iron-peak elements seen in SN1987A (\cite{mcc93}; \cite{tue90}).

Early during the explosion, while most of the shocked matter is moving
outward, some of it is still falling inward. There is simultaneously
explosion and accretion. The cooler material at 90$^{\circ}$ in Figure 27 is
plunging inward even 40 milliseconds after the explosion is ignited, though by
this time very little additional matter is sinking and the general flow is
explosive. The continuous accretion early during the explosion is not
necessary to power the blast wave, and subsides after $\sim$70 milliseconds
without any effect on its viability. Specifically, there are no accretion
fingers
that dive into the core after $\sim$70 milliseconds and the wind is clearing
out the
interior. However, this ``broken symmetry'' not
seen in the 1-D simulations may be the means by which nature fattens the
core to observed neutron star masses $(M_g\sim 1.35\ \mdot, M_B\sim 1.5\
\mdot)$,
while exploding quickly within 100 milliseconds of bounce. An ``early'' 1-D
explosion of a 15 $\mdot$ star leaves behind too little mass (Figure
7). However, at $t=408$ milliseconds into the 2-D {\bf star} simulation, the
mass left behind is still only 1.32 $\mdot$.

It is into the channels created by the rising plumes that the protoneutron
star wind material flows. This anisotropic wind continues to drive the inner
explosion. Two hundred milliseconds after bounce, a clump in the wind has an
entropy near 60 (Figure 27), the highest in the flow and higher than achieved
in 1-D (Figure 14). Such high entropies are achieved because wind material
emerging from the core is slowed down temporarily by the more slowly moving
material it encounters further out.  This allows such matter to be heated
longer by the core neutrinos and thereby to achieve higher entropies.
How these hot spots evolve has yet to be determined and, as
stated previously, may be of relevance to the r-process (\cite{tak94};
\cite{wo94a}). The hot spot in Figure 27 has a mass near $10^{-5\pm 1}\mdot$,
depending on how its third dimension is treated.

At $t=408$ milliseconds, the region interior to 120 kilometers that was
vigorously unstable before explosion is now host to a stable, spherical wind,
though
there are still interacting non-radial streams in the region between 150 and
600 kilometers.  By 348 milliseconds, as Figure 25 depicts, the shock has
reached
$\sim$700 kilometers and the inner edge of the oxygen zone that has fallen in
to meet it. The post-shock temperatures are $\sim$0.8 to 0.9 MeV and the
post-shock densities are $\sim 10^8$ gm/cm$^3$. The immediate post-shock
entropies are between 8 and 11 and $Y_e$ equals $\sim 0.5$. In material with
these thermodynamic characteristics, alpha particles are $\sim 90$\% by mass,
and the rest is neutrons. However, in the high-entropy, ``low'' density plumes
depicted in Figures 25--27, the alphas are dissociated into neutrons and
protons. It is on the free neutrons and protons that the neutrinos from the
core can be absorbed. Therefore, despite the fact that the matter is at
progressively larger radii, the high entropy fingers continue to be heated,
albeit at a decreasing rate. Nevertheless, from 318 to 408 milliseconds, the
entropy of the leading edges of the 60$^{\circ}$ plume increases by no more
than a few units, while that of the material at the base of the 60$^{\circ}$
plume (at smaller radii) is increased by as much as ten units (and starts from
higher values).

A major concern of our {\bf star} results is the electron fraction of some of
its ejecta. Most of the exploding material depicted in Figure 28 has an
unoffending $Y_e$ near 0.5. However, the most vigorous and obvious plume at
60$^{\circ}$ from the vertical in Figure 29 has a $Y_e$ in most of its mass of
0.43 to 0.46. What is more, $Y_e$ approaches 0.4 near its base in the inner
wind-fed region. The mass of the neutron-rich plume may total $10^{-3}\mdot$,
which may be too much to be consistent with the observed abundances of
neutron-rich isotopes near the iron-peak. Since the ejected $Y_e$ is modified
by the electron- and anti-electron-neutrino fluxes, the solution to this
dilemma may simply be further evolution or better neutrino physics
(\cite{ful94}). It may
also be solved by a longer delay to explosion, since the accreting matter is
gradually thinning out with time and the post-shock electron capture rates are
decreasing. This has the effect of ejecting less mass, and with higher
$Y_e$'s.

At the end of the calculation at $t=408$ milliseconds, the leading edge of the
shock is at 2200 kilometers. The matter temperatures between a radius of 1000
kilometers and the shock are between 0.45 and 0.6 MeV at all angles. Though
the final explosion energy has not yet saturated,
the kinetic energy is $0.42\times 10^{51}$ ergs, the internal energy interior
to the shock is $0.06\times 10^{51}$ ergs, and the dissociation energy is
$0.38\times 10^{51}$ ergs, two hundred milliseconds after bounce.
The total energy is rising at an instantaneous rate of $4\times 10^{51}$
ergs/s.
\vskip8truept
\centerline{{\bf c. Pulsar Kicks}}
\smallskip
The convective fluctuations in mass accretion flux and neutrino luminosity
before and during
explosion suggest a natural mechanism for imparting kicks to neutron
stars. The observed average kick speed of radio pulsars is $\sim$450 km/s
(\cite{lyn94}). The inner 15 kilometers of the {\bf star} model was
pinned and done in 1-D and a calculation with an angular range of 360$^{\circ}$
(or 180$^{\circ}$) is called for. However, with the fluctuating momentum in
the $y$ direction $({\rm at\ }\theta=0^{\circ})$, we can estimate the recoil
velocities that would have been imparted to the core.
The recoil speed in the $y$ direction
fluctuated on a short convective timescale of 3 milliseconds, but grew on
average to peak at 295 milliseconds $(\sim15$ milliseconds before explosion)
at a value of $\sim$180 km/s.  Convection deep in the potential well causes
the core to shake significantly. The recoil due
to the integrated anisotropic neutrino emissions before explosion is smaller
than that due to the mass motions before explosion by a factor of roughly
${v\over c}$, as can be demonstrated simply by noting that the accretion
$L_\nu$ is
proportional to ${v^2\over 2}$, while the recoil due to mass accretion scales
as $v$, where $v$ is the post-shock accretion  speed. If we multiply this peak
one-dimensional speed of 180 km/s by $\sqrt3$ to account for three dimensions,
we obtain a peak recoil speed of 300 km/s due merely to the oscillations of
the core
near explosion. This estimate does not include the effects of mass motions
at the polar caps omitted in the {\bf star} calculation.
Note that the symmetry of our {\bf star} calculation does not allow us to
estimate the precise magnitude of this effect with any confidence. However, we
think that we have in the vigorous convective motions of the
material between the stalled shock and the neutrinospheres a natural mechanism
for imparting to many young pulsars high proper motions. This is
accomplished without the use of magnetic fields. However, velocities above
500 km/s may still be difficult to explain, though we surmise that the
peak vigor of the convection is larger for the fatter cores associated with
more massive progenitors. This would suggest that the RMS kick speed of a
neutron star may increase with progenitor mass, all else
being equal. Very mild support for this hypothesis comes from the smaller
proper motion observed for the Crab pulsar ($v_t\sim 150$ km/s;
\cite{har93}), which \refmark{Nomoto \etal\ (1982)}{nom82} conclude originated
from a smaller 8--10 $\mdot$ progenitor.  However, for a perfectly spherical
progenitor, a long delay to explosion would cancel the asymptotic recoil speed
convection could impart (\cite{jan94}), since the initial total momentum would
be
zero and the mass in the shocked envelope is gradually decreasing with
time. For a given progenitor, there is a critical explosion time for peak
pulsar recoil, after which the convective effect we have outlined
diminishes. What the actual explosion and recoil systematics are has yet to be
determined.

A related and intriguing effect of neutrino-driven convection is its tendency
to impart to the residue not only {\it linear} momentum, but {\it angular}
momentum as well. The g-mode
modulated accretion plumes strike at a variety of angles and torque the core
stochastically. The numbers of our {\bf star} simulation suggest that the
core is easily left with {\bf rotation} periods of less than one
second. Thus,
even if the Chandrasekhar core of a massive star is not rotating at collapse,
the neutron star residue should be left rotating with respectable pulsar
periods. If the kicks and periods are convection-induced and the white-dwarf
cores are rotating very slowly at collapse, we
would expect a roughly linear correlation between the rotation frequencies of
pulsars at birth and their proper motions $(\omega\sim{\Delta v\over R}$,
where $R\sim30$ kilometers). We would also expect that the kick
direction and spin axis would be very approximately at right angles.  However,
any significant initial angular momentum just before collapse might easily
confuse and suppress this correlation, as might a significantly aspherical
core (cf. \refmark{Bazan and Arnett (1994)}{baz94}). In addition, any
systematic trend of
initial core rotation period with progenitor mass would have to be understood
(as would the effect of rotation itself on neutrino-driven convection) to
extract this convective correlation unambiguously. Nevertheless, it is
clear that a core need not be rotating before collapse for a pulsar to be born
with an interesting period.

\section{Conclusions}
Our 2-D calculations imply that the breaking of spherical symmetry may be
central
to the supernova phenomenon. The explosion does not erupt spherically, but in
crooked fingers and is fundamentally heterogeneous. This conclusion is
consistent with the nickel bullets and the ragged and skewed infrared iron,
nickel, and cobalt line profiles observed in SN1987A, the ``shrapnel''
observed in CAS A, N132D, and the Vela supernova remnant (\cite{asc94}), and
the large observed pulsar proper motions. In this paper, we have attempted to
understand and extend the current theory of supernova explosions. To do so, we
have performed new one- and two-dimensional radiation/hydrodynamic simulations
of all phenomena from collapse through explosion. The 1-D calculations allowed
us to analyze the context in which the 1-D neutrino-driven mechanism should be
understood and to highlight its limitations. The role of progenitor structure
on the 1-D development of collapse was explored. It was shown that
the two classes of core structure, compact and extended, can evolve in
quantitatively, and perhaps qualitatively, different ways.

With our 2-D {\bf star} simulation of the core of a 15 $\mdot$ progenitor, we
have verified the potential importance of neutrino-driven convection in
igniting a supernova explosion (\cite{bet90} and \refmark{HBC}{hbc}). After a
delay of $\sim$100 milliseconds, the core exploded aspherically driven by
rising neutrino-heated bubbles that developed after another 100 milliseconds
into tangled fingers. These fingers seem to be generic features of supernova
explosions.
We do not see a cascade to a dominant
``$\ell=1$'' mode,  either before or after explosion, and both small- and
large-scale structures are always visible. Before
explosion, vigorous columnar downflows develop and dance over the
neutrinosphere, but they break up within at most 20 milliseconds. The flow
fluctuates between ordered and ``chaotic'' motion and modes of $\ell$ equal 2
to 10 are generally in evidence. Internal gravity waves (g-modes) are excited
by the downflows and easily achieve non-linear amplitudes of 5 to 10
kilometers. These g-modes feed back onto the downflows and modulate both the
convective motions between the shock and the neutrinospheres and the positions
of the downflowing plumes. The convective motions get progressively more
vigorous with time. The rising bubble speeds reach values of 30,000 km/s
within 50 milliseconds. These bubbles hit the shock and result in oscillations
with time and variations with angle of its radius that can be 30\%. The
convective zone can overshoot by as much as ten kilometers and the consequent
enhancement in neutrino luminosity can be 10--20\%. Lepton-driven convection
beneath the neutrinospheres starts early and is maintained even after
explosion. Its boosting effect on the neutrino luminosities that drive the
convection is persistent, but seems swamped by the variations with angle and
time of the fluctuating accretion luminosity. The latter varies on a
convective timescale of $\sim$3--5 milliseconds and by as much as a factor
of three in angle. The angle-averaged luminosities vary by as much as 60\%
(Figure 16b).

As in 1-D, the explosion appears to be a critical phenomenon
with a critical condition.
In 2-D, the rising balloons, upon
encountering the shock, are immediately advected inward by the powerful mass
accretion flux raining down. They do not dwell near the shock. In fact, the
net mass flux through the shock is approximately equal to the mass flux onto
the core and mass does not accumulate in the convective zone. The mass between
the shock and the neutrinospheres {\it decreased} by about a factor of three
in the the {\bf star} calculation during the pre-explosion boiling
phase
that lasted $\sim$100 milliseconds ($\sim$30 convective turnover times). {\it
All} the matter that participates in the ``convection'' before explosion
eventually leaves the convection zone and settles onto the core. A given
parcel of matter may ``cycle'' one or two times before settling inward (and a
large fraction never rises), but more than three times is rare. The boiling
zone is resupplied with mass by mass accretion through the shock and a
secularly evolving steady-state is reached. This steady-state is similar to
that achieved in 1-D, but, due to the higher dwell time in the gain region, the
average entropy in
the 2-D envelope is larger and its entropy gradient is flatter. These effects,
together with the dynamical pressure of the buoyant plumes, serve to increase
the steady-state shock radius over its value in 1-D by 30\%--100\%. It is this
effect of boiling that is central to its role in triggering the explosion, for
it thereby lowers the critical luminosity threshold.  The
lowering of the effective critical curve (Burrows and Goshy 1993) allows the
actual model trajectory in
$L_{\nu_e}$ vs. $\bfdot M$ space to intersect the new critical curve.
The physical reasons
for the lowered threshold are straightforward: a large $R_s$ enlarges the
volume of the gain region, puts shocked matter lower in the gravitational
potential well, and lowers the accretion ram pressure at the shock for a given
$\bfdot M$. Since the ``escape'' temperature $(T_{\rm esc})$
decreases with radius faster than the actual matter temperature $(T)$ behind
the shock, a larger $R_s$ puts a larger fraction of the shocked material above
its local escape temperature. $T>T_{\rm esc}$ is the condition for
a thermally-driven corona to lift off of a star.
Since supernovae are driven by neutrino heating, they are coronal
phenomena, akin to thermally-driven winds, though initially bounded by an
accretion
tamp. Neutrino radiation pressure is unimportant.

We see no evidence that convection in and of itself leads to a
``building'' or accumulation of energy before
explosion. In fact, in the {\bf star} calculation, the total energy in the
overturning
region actually {\it decreased} monotonically before explosion.  The kinetic
energy
alone increased.
If almost anytime before explosion the accretion
rate was not allowed to decay further, but was frozen, the convection
would boil in a steady state and the star
would not explode. Furthermore, we find that the accuracy and specifics of the
neutrino scheme employed can make a qualitative difference in the outcome.
We conclude that even with Bethe convection, the explosion is not
assured and the outcome is still a quantitative question requiring further
exploration.

After explosion, the ejecta emerge in many contorted fingers in which the
density can vary with angle by more than a factor of four. Early during the 2-D
explosion,
unlike in the 1-D calculations, some matter is still accreting in long, thin,
bent, lower-entropy plumes. Later, all of the matter is exploding, though the
entropy in the outer ejecta varies with angle from 10 to 40. However, as late
as 70
milliseconds after explosion, some matter is still crashing down to within 200
kilometers
of the
core. By 100 milliseconds, a stable, neutrino-driven wind is clearly being
blown from
the protoneutron star. This wind is partially channelled by the low-density
fingers and
helps to drive the explosion. It is roughly spherical, but roils the matter it
encounters 200 to 600 kilometers away (see also Figure 13) and does not allow
matter to
penetrate onto the core. The entropy in hot
spots at the base of the plumes and in the wind has reached 60, 100
milliseconds into the explosion. What its entropy is asymptotically (the
r-process?) remains to be seen.

The kinetic energy of the explosion at 408 milliseconds is $0.42\times
10^{51}$ ergs, the internal energy interior to the shock is $0.06\times
10^{51}$ ergs, and the stored dissociation energy is $0.38\times10^{51}$
ergs. At the end of the
calculation, the kinetic energy is growing at a rate
of $2.2\times 10^{51}$ ergs/s and the total explosion energy is growing at
twice that rate. The supernova energy will be the sum of the above with the
neutrino energy yet to be deposited, the gravitational energy of the ejecta,
the binding energy of the unshocked mantle, and the thermonuclear
component. The scale of a supernova's explosion energy may ultimately be set by
the initial binding energy of the progenitor envelope (Figure 12). It is very
important to note that in the {\bf star} calculation, and perhaps
generically in supernovae, energy deposition is not instantaneous. Even after
the inner edge of the oxygen zone is encountered by the shock, neutrinos
continue to heat the ejecta. The total supernova energy is determined only
after the shock has propagated many thousands of kilometers. This means among
other things that the post-shock temperatures are lower during explosive
nucleosynthesis than heretofore assumed and as a consequence that the
explosive yield of $^{56}$Ni is lower. This may help solve the
$^{56}$Ni overproduction problem alluded to in \S IV.

We see evidence that pulsar kick velocities of hundreds of kilometers per
second can be imparted hydrodynamically to the core by the chaotic convective
motions before
and during explosion. However, the precise magnitude and systematics with
progenitor of
these recoils remain to be determined. In addition, due solely to
neutrino-driven convection, a core can be left with a rotational period less
than one second (even if it is not rotating at collapse). These intriguing
possibilities of direct relevance to pulsar properties will be explored with
future calculations, which will ultimately have to be done in 3-D.

Two major concerns of the {\bf star} model are the low baryon mass of its
residue and the large amount of neutron-rich material ejected (perhaps
$10^{-3}\ \mdot$). The latter may have embarrassing nucleosynthetic
consequences, but may be solved if neutrino capture on nuclei is
included in the calculation (\cite{ful94}).  However, both problems might be
solved if the delay to explosion is longer. The time between bounce and
explosion is a quantitative question
that rests in no small part on the accuracy of the neutrino algorithm and the
progenitor structure. A major index of a supernova calculation is the
steady-state shock radius it achieves. Variations with progenitor and among
modelers of the hydrodynamics and outcomes might be traced to its value, which
ranges in the theories between 60 and 150 kilometers.

Figure 16b depicts the average neutrino luminosities versus time for the {\bf
star} calculation and its 1-D analog. Apart from demonstrating the large
fluctuations alluded to previously and the growing effect of overshoot before
explosion, it shows that the average luminosities drop by about a factor of
two over a period of $\sim$10--15 milliseconds right at the explosion. The
detection of this explosion signature by underground neutrino telescopes may
be feasible and would shed needed light on supernova dynamics (\cite{bkg92}).

The calculations summarized in this paper are but the beginning of a long
series of numerical explorations by us and by others to understand the
multi-dimensional aspects of supernova explosions. The systematics with
progenitors, the possibility of dynamo action (\cite{tom93}), the recoil
speeds, the gravitational radiation signatures, the effects of neutrino
viscosity (\cite{bl88}), the formation of black holes, and the nucleosynthetic
yields have yet to be adequately derived or investigated. Furthermore, the
influence of resolution, better neutrino transport, and the third dimension
have not been clarified. Our
experience with this extensive set of numerical simulations convinces us that
most future issues of supernova theory should be pursued with
multi-dimensional tools.

\vfill\eject

\section{Acknowledgements}
Conversations with Grant Bazan, Willy Benz, Chris Fryer, Stirling Colgate,
Thomas Janka, Stan Woosley, and Marc Herant are gratefully acknowledged.
Special thanks are extended to the referee, Ewald M\"uller, for his many useful
suggestions that materially improved the manuscript.
The calculations were performed on the Cray Y/MP-8 at the NCSA in Illinois, on
the Cray Y/MP-16 at the Pittsburgh Supercomputer Center in Pennsylvania, and
on the Cray Y/MP-8 at the San Diego Supercomputer Center in
California.
Post-processing and code development were performed on two SGI
4D/35's and on an INDIGO2 XZ.
For access to the SDSC, we wish to thank
Dave DeYoung, and for continued access to the other machines we thank Cathy
Milligan
and Vicki Halberstadt. The authors express appreciation to Tom Weaver and Stan
Woosley for making a subset of their progenitor models available in
machine-readable form. This work was supported by the NSF under grant
AST92-17322 and by NASA under grant NAGW-2145.
The color
figures in this paper and mpeg-format movies depicting aspects
of the explosion can be found at WWW address
http://lepton.physics.arizona.edu:8000/.


\newpage

\centerline{{\bf Figure Captions}}

\noindent Figure 1a: Logarithm of the density vs. interior mass
for selected times during the evolution of model sequence w15t.
The curves labeled ``a'' through ``d'' correspond to times of
0.0200, 0.2075, 0.2106, and 0.2691 seconds, respectively, where
time is measured from the beginning of the calculation.  Core
bounce occurred near t = 0.209 seconds.

\noindent Figure 1b: Logarithmic density profiles as in Fig. 1a;
but for model sequence w20t.  The curves labeled ``a'' through
``d'' are at times t = 0.0200, 0.5465, 0.5594, and 0.6020 seconds,
respectively.  Core bounce occurred near t = 0.546 seconds.

\noindent Figure 2a: Profiles of velocity vs. mass for model
sequence w15t.  The labeled curves (a-d) are at times of 0.0200,
0.2091, 0.2106, and 0.2691 seconds, respectively.

\noindent Figure 2b: Profiles of velocity vs.\ mass for model sequence
w20t.  The labeled curves (a-d) are at time of 0.0200, 0.5472,
0.5500, and 0.6020 seconds, respectively.

\noindent Figure 3a: Profiles of entropy (per Boltzmann's constant
per baryon) vs. mass for model sequence w15t.  The labeled curves
(a-d) are at times of 0.0200, 0.2106, 0.2187, and 0.2691 seconds,
respectively.

\noindent Figure 3b: Profiles of entropy
vs. mass for model sequence w20t.  The labeled curves
(a-d) are at times of 0.0200, 0.5486, 0.5700, and 0.6020 seconds,
respectively.

\noindent Figure 4a: Profiles of electron fraction vs. mass for
model sequence w15t.  The labeled curves
(a-d) are at times of 0.0200, 0.2087, 0.2187, and 0.2691 seconds,
respectively.

\noindent Figure 4b: Profiles of electron fraction vs. mass for
model sequence w20t.  The labeled curves
(a-d) are at times of 0.0200, 0.5465, 0.5551, and 0.6020 seconds,
respectively.

\noindent Figure 5a: Plots of the shock radius and neutrinosphere
radii vs. time for model sequence w15t.  The zero point of time
is taken to be the time of core bounce, t = 209 ms.

\noindent Figure 5b: Plots of the shock radius and neutrinosphere
radii vs. time for model sequence w20t.  The zero point of time
is taken to be the time of core bounce, t = 547 ms.

\noindent Figure 5c: Plots of the shock radius
vs. time for the 2-D {\bf star} model and its 1-D analog.  The zero point of
time
is taken to be the beginning of the calculation.  This plot illustrates one of
the
major multi-dimensional effects.

\noindent Figure 6a: Plots of the neutrino luminosities vs. time
for model sequence w15t (One ``foe'' is $10^{51}$ ergs.).  The zero point of
time is that of core
bounce, as in Fig. 5a.

\noindent Figure 6b: Plots of the neutrino luminosities vs. time
for model sequence w20t (One ``foe'' is $10^{51}$ ergs.).  The zero point of
time is that of core
bounce, as in Fig. 5b.

\noindent Figure 7: Mass accretion rate through the shock and the total mass
interior to
shock vs. time for model sequences w15t and w20t.  The zero point
of time is that of core bounce in all cases.

\noindent Figure 8: Profiles of velocity vs. radius for model
sequence w15n.  The total evolution time is approximately 429 ms.

\noindent Figure 9: The total kinetic energy vs. time from core
bounce for model sequences w15n and w15e3.

\noindent Figure 10: Plots of iron core mass vs. ZAMS mass for
models from Weaver and Woosley (1995, ww) and Nomoto and Hashimoto (1988, NH)
(see text).  Note the large spread in iron core mass (several tenths of a
solar mass) for a given total mass. wwsb is the preferred model of ww, while
wwna with severly reduced semi-convection is disfavored.

\noindent Figure 11: Profiles of density vs. mass for the inner
2 M$_{\odot}$ of five different initial models from the Weaver
and Woosley (1995) ``sb'' series. Locations of the iron core edges are
indicated by the open circles.

\noindent Figure 12: Profiles of the binding energy vs. mass for
four of the Weaver and Woosley (1995) progenitor models.  Locations
of the iron core edges are indicated by open circles.

\noindent Figure 13: Profiles of velocity vs. radius for the model
sequence w15e3.  Note the prominent second shock developing at
approximately 800 km, driven by neutrino heating from below and the
steady neutron star wind.

\noindent Figure 14: Entropy vs.\ radius for the final timestep in the w15e3
model sequence. The peak at 800 km is due to the shock at the interface
between the neutron star wind and the more slowly moving material beyond
it. The
peak at 1800 km is that which was created at core bounce, and the peak near
2400 km was created when the bounce shock encountered the inner edge of the
silicon shell.

\noindent Figure 15: The $Y_e$ distribution at $t=238$ milliseconds $(\sim30$
milliseconds after bounce). The figure is 150 km$\times$150 km. The dominant
flows are still inward and the degree of overshoot is significant. Velocity
vectors are superposed.

\noindent Figure 16a: Neutrino luminosities vs.\ angle at three different
times in the {\bf star} model sequence. The luminosity is defined to be the
flux along the indicated angular coordinate times the total area at
infinity. The times are 0.2502, 0.2703, and 0.2977, seconds; the
first time is shortly after the onset of the instability, and the final time
is immediately prior to the explosion. The curves for the final time are
labeled with ``e.''

\noindent Figure 16b: The three neutrino luminosities vs.\ time for the {\bf
star} model
sequence and its 1-D analog. The luminosities plotted for
the 2-D {\bf star} model are the angle-averaged quantities.

\noindent Figure 17: Same as Figure 15, but for 270 milliseconds. Multiple
plume and bubble structures can be seen.

\noindent Figure 18: The $Y_e$ distribution at $t=299$ milliseconds, $\sim10$
milliseconds before explosion. The plot is 150 km on a side and velocity
vectors trace the flow. Note the vigorous bubble at $\sim20^{\circ}$ that has
pushed the shock out to $\sim135$ kilometers and the corrugations at the base
of the convective zone.

\noindent Figure 19: The $Y_e$ distribution with velocity vectors at $t=304$
milliseconds, $\sim5$ milliseconds before explosion. The scale is as in Figure
15. The size of the convective zone has increased significantly since bounce
(or since in Figure 15). A strong downward plume can be seen at
$\sim60^{\circ}$ penetrating deep below the neutrinospheres.

\noindent Figure 20: The $Y_e$ distribution for the {\bf star} calculation at
$t=307$ milliseconds, just before explosion. The scale is 150 km$\times$150
km. The material near $\theta=0^{\circ}$ is moving inward and the chaotic
motions are quite vigorous and varied (see the velocity vectors that follow
the flow).

\noindent Figure 21: The $Y_e$ distribution at $t=311$ milliseconds
calculation, within milliseconds of exploding. Some material is still falling
in. The scale is as in Figure 20 and velocity vectors indicate the direction
and magnitude of the motions.

\noindent Figure 22: Same as Figure 19, but the entropy distribution (at
$t=304$ milliseconds) without velocity vectors. The hot spots in blue have
reached entropies of $\sim32$ units.

\noindent Figure 23: Neutrinosphere radii at the times indicated in Figure
16a. For each neutrino species, the uppermost curve corresponds to the first
time indicated (0.2502 sec), and the trend with time is uniformly
downward. Thus the lowest curves on this plot correspond to the curves labeled
``e'' in Figure 16a.

\noindent Figure 24: The entropy distribution at $t=318$ milliseconds,
($\sim10$ milliseconds into the explosion). The scale is 1500 km$\times$1500
km. The hot blue bubbles driving the explosion are clearly seen.

\noindent Figure 25: Same as Figure 24, but at $t=348$ milliseconds. The
shock is encountering the inner edge of the oxygen zone near 700 kilometers.

\noindent Figure 26: Same as Figure 24, but at $t=378$ milliseconds. The
highest entropies (at $\sim38$ units) are found in clumps.

\noindent Figure 27: The entropy distribution at $t=408$ milliseconds
($\sim$100 milliseconds into the explosion) for the {\bf star} model. The
horizontal and vertical scales are 2500 kilometers. Note the hot spots near
$\theta=45^{\circ}$, where the entropy is $\sim60$ units. Note also the warm
fingers (with mushroom caps) and the thin cooler tendrils.

\noindent Figure 28: Same as Figure 27 (at $t=408$ milliseconds), but for
$Y_e$, with velocity vectors superposed. The tongue of material with $Y_e's$
of 0.43--0.46 is in evidence, as is a slower plume near $\theta=20^{\circ}$.

\noindent Figure 29: Same as Figure 27, but for the radial velocity. The
``jets'' in places are moving faster than $5\times 10^9$ cm/s ($\equiv$ 50,000
km/s).

\noindent Figure 30: The same as Figure 27 ($t=408$ milliseconds), but for
$\log_{10}\rho$. The blue dots in the most vigorous finger have densities
below $10^6$ gm/cm$^3$, and entropies near 35. Their densities are below those
immediately in front of the shock.

\newdimen\digitwidth
\setbox0=\hbox{\rm0}
\digitwidth=\wd0
\catcode`?=\active
\def?{\kern\digitwidth}
\begin{table*}[h]
\begin{center}
Table 1
\end{center}
\begin{center}
Central Density Rise Times for 1-D Models
\end{center}
\begin{center}
\begin{tabular}{rlllll} \hline \hline
& \multicolumn{5}{c}{Model Sequence} \\ \cline{2-6}
	             & w15t   & w15e3  & w15n  & w20t  & w20n \\ \hline
*$\Delta$t 10-11 (ms) & 167    & 167    & 186    & 145   & 192 \\
$\Delta$t 11-12 (ms) & ?25.1  & ?25.1  & ?35.9  & ?20   & ?32.5\\
$\Delta$t 12-13 (ms) & ??5.25 & ??5.25 & ??8.5  & ??4.75& ??8.48 \\
$\Delta$t 13-14 (ms) & ??1.53 & ??1.53 & ??2.4  & ??1.30& ??2.51 \\ \hline
\multicolumn{6}{l}{*Time for the central density to go from $10^{10}$ to
$10^{11}$ gm/cm$^3$,}\\
\multicolumn{6}{l}{\phantom{*}other rows correspondingly.}\\
\end{tabular}
\end{center}
\end{table*}
\begin{table*}[h]
\begin{center}
Table 2
\end{center}
\begin{center}
Evolution of Fe and Si Edges for 1-D Models
\end{center}
\begin{center}
\begin{tabular}{lccccc} \hline \hline
& \multicolumn{5}{c}{Model Sequence} \\ \cline{2-6}
              & w15t  & w15e3                & w15n        &w20t  & w20n\\
\hline
R(Fe) at t=0 (cm)     & 1.15 $\times 10^{8}$ & 1.15 $\times 10^{8}$ & 1.15
$\times 10^{8}$ &
2.21 $\times 10^{8}$  & 2.21 $\times 10^{8}$ \\
R(Fe) at Bounce (cm)  & 5.38 $\times 10^{7}$ & 5.38 $\times 10^{7}$ & 1.89
$\times 10^{7}$ &
1.43 $\times 10^{8}$  & 8.74 $\times 10^{7}$ \\
R(Si) at t=0 (cm)     & 2.46 $\times 10^{8}$ & 2.46 $\times 10^{8}$ & 2.46
$\times 10^{8}$ &
3.27 $\times 10^{8}$  & 3.27 $\times 10^{8}$ \\
R(Si) at Bounce (cm)  & 1.71 $\times 10^{8}$ & 1.71 $\times 10^{8}$ & 1.71
$\times 10^{8}$ &
2.54 $\times 10^{8}$  & 2.13 $\times 10^{8}$ \\
R(Si) at Finish (cm)  & 1.61 $\times 10^{8}$ & 3.45 $\times 10^{8}$ & 3.63
$\times 10^{8}$ &
2.41 $\times 10^{8}$  & 9.51 $\times 10^{7}$ \\
Time at Bounce (ms)   & 208                  & 209            &  244 &
547                   & 744 \\
\hskip-3pt *t$_{\rm sh}$(Fe) (ms) & 233 (24.7) & 232 (23.3)   & 245 (1.93)
     &
$\ldots$              & 769 (25.3) \\
\hskip-3pt *t$_{\rm sh}$(Si) (ms) & $\ldots$ & 302 (93.0)     & 298 (54.3) &
$\ldots$              &  $\ldots$ \\
Time at Finish (ms)   & 270 (61.7)           & 401 (192)      &  432 (188) &
602 (55.3)            & 905 (161) \\ \hline
\multicolumn{6}{l}{*t$_{\rm sh}$ is the time the shock encounters the
composition boundary.}\\
\multicolumn{6}{l}{\phantom{*}(numbers in parentheses are times after bounce in
milliseconds.)}\\
\end{tabular}
\end{center}
\end{table*}
\begin{table*}[h]
\begin{center}
Table 3
\end{center}
\begin{center}
Index of Color Figures
\end{center}
\begin{center}
\begin{tabular}{lllcc} \hline \hline
	&		&	&		&\\
Figure \#&File Name	&Epoch(ms)&Scale (km$\times$km) &Variable\\
\hline
\multicolumn{5}{c}{\underbar{Before Explosion}} \\
15	&starrcx	&238	&150$\times$150	&$Y_e$\\
17	&starrdm	&270	&150$\times$150	&$Y_e$\\
18	&starree	&299	&150$\times$150	&$Y_e$\\
19	&starrei	&304	&150$\times$150	&$Y_e$\\
20	&starrek	&307	&150$\times$150	&$Y_e$\\
21	&starren	&311	&150$\times$150	&$Y_e$\\
22	&starrei	&304	&150$\times$150	&Entropy\\
\multicolumn{5}{c}{\underbar{After Explosion}}\\
24	&starres	&318	&1500$\times$1500	&Entropy\\
25	&starrfm	&348	&1500$\times$1500	&Entropy\\
26	&starrgf	&378	&1500$\times$1500	&Entropy\\
27	&starrgy	&408	&2500$\times$2500	&Entropy\\
28	&starrgy	&408	&2500$\times$2500	&$Y_e$\\
29	&starrgy	&408	&2500$\times$2500	&$v_r$\\
30	&starrgy	&408	&2500$\times$2500	&$\log \rho$\\
\hline
\end{tabular}
\end{center}
\end{table*}
\eject

\end{document}